\documentclass[nofootinbib,aps,twocolumn]{revtex4}
\usepackage{graphicx}
\usepackage{amsfonts}
\usepackage{amssymb}
\usepackage{amsbsy}
\usepackage{amsmath}
\usepackage{mathrsfs}
\usepackage{latexsym}
\usepackage{natbib}
\usepackage{bm}
\usepackage{subfigure} 
\usepackage{graphicx}
\usepackage{color}
\usepackage{wasysym}
\usepackage{mathbbol}
\usepackage{bigints}
\usepackage{float}
\usepackage{booktabs}

\newcommand{\intinf}{\int_{-\infty}^{\infty}} 
\newcommand{\intsinf}{\int_{0}^{\infty}} 
 
\newcommand{\no}{\nonumber} 
\newcommand{\p}{\prime}

\begin{document}

\title{Entanglement harvesting between two inertial Unruh-DeWitt detectors from non-vacuum quantum 
fluctuations}

\author{Dipankar Barman}
\email{dipankar1998@iitg.ac.in}

\author{Subhajit Barman}
\email{subhajit.b@iitg.ac.in}

\author{Bibhas Ranjan Majhi}
\email{bibhas.majhi@iitg.ac.in}

\affiliation{Department of Physics, Indian Institute of Technology Guwahati, 
Guwahati 781039, Assam, India}


\date{\today}

\begin{abstract}

Entanglement harvesting from the quantum field is a well-known fact that, in recent times, is being 
rigorously investigated further in flat and different curved backgrounds. The usually understood 
formulation studies the possibility of two uncorrelated Unruh-DeWitt detectors getting entangled 
over time due to the effects of quantum vacuum fluctuations. Our current work presents a thorough 
formulation to realize the entanglement harvesting from non-vacuum background fluctuations. In 
particular, we further consider {\it single excitation} field states and a pair of {\it inertial 
detectors}, respectively, in $(1+1)$ and $(1+3)$ dimensions for this investigation. Our main 
observation asserts that {\it entanglement harvesting is suppressed compared to the vacuum 
fluctuations} in this situation. Our other observations confirm a non-zero individual detector 
transition probability in this background and vanishing entanglement harvesting for parallel 
co-moving detectors. We look into the characteristics of the harvested entanglement and discuss its 
dependence on different system parameters.

\end{abstract}

\maketitle

\section{Introduction}\label{Introduction}

The investigations to understand quantum entanglement-related phenomena for relativistic particles 
have gained significant momentum recently (see \cite{FuentesSchuller:2004xp, Reznik:2002fz, 
Lin:2010zzb, Ball:2005xa, Cliche:2009fma, MartinMartinez:2012sg, Salton:2014jaa, 
Martin-Martinez:2015qwa, Cai:2018xuo, Menezes:2017oeb, Menezes:2017rby, Zhou:2017axh, 
Floreanini:2004, Pan:2020tzf, Barman:2021oum,Kane:2021rhg,Barman:2021igh,Barman:2022utm}). This veneration of 
quantum entanglement lies in its fascinating nature to be able to distinguish a quantum phenomenon 
from a classical one. Furthermore, the discovery of gravitational waves provided additional interest 
in this direction. Though predicted entirely from the classical considerations, these discoveries 
encourage one to look also for the signatures of the quantum phenomena from the early universe. 
Therefore, entanglement is revered as an arena to broaden our conceptual horizon, and it is also 
relevant from many practical fronts. In particular, it finds its applications in many forerunning 
expectations of modern technology like quantum communication, cryptography, and teleportation 
\cite{Tittel:1998ja, Salart-2008}. The experimental verification of entanglement in systems with 
photons and even in electrons \cite{PhysRevA.87.053822, Hensen:2015ccp} makes these possibilities 
all the more lucrative. In this regard, an enormous amount of interest is given to the possibility 
of two initially   uncorrelated systems getting entangled over time, which is better known in the 
literature as the {\it entanglement harvesting}. It predicts that from the vacuum fluctuation of the 
quantum fields, depending upon the motions of the systems, one can harvest entanglement between two 
or more systems when they are interacting with the background quantum fields. Moreover, one can 
further utilize this for quantum information-related works like quantum teleportation 
\cite{Hotta:2008uk, Hotta:2009, Frey:2014}. The works \cite{VALENTINI1991321, Reznik:2002fz, 
Reznik:2003mnx, Salton:2014jaa} first introduced the idea of entanglement harvesting. In particular, 
the work \cite{Reznik:2002fz}, where Reznik pioneered this phenomenon, considers point-like 
two-level Unruh-DeWitt (UD) hypothetical particle detectors. In this work, the author provided the 
understanding of entanglement harvesting between two accelerated and causally disconnected UD 
detectors (initially conceptualized to understand the Unruh effect \cite{Unruh:1976db, 
Unruh:1983ms}), which interact with the background massless scalar field. Since these articles, 
there has been a significant amount of research towards the realization of entanglement harvesting 
from the quantum field vacuum in different spacetime backgrounds as well as for various types of 
motions of detectors \cite{Martin-Martinez:2013eda, Lorek:2014dwa, VerSteeg:2007xs, Brown:2014pda, 
Pozas-Kerstjens:2015gta, Pozas-Kerstjens:2016rsh, Martin-Martinez:2015qwa, Kukita:2017etu, 
Sachs:2017exo, Trevison:2018ear, Li:2018xil, Henderson:2017yuv, Henderson:2020ucx, 
Stritzelberger:2020hde, Koga:2018the, Ng:2018ilp, Koga:2019fqh, Brown:2013kia, Simidzija:2018ddw, 
Lima:2020czr, Barman:2021bbw, Barman:2021kwg}. Also the relative motion of the detectors has 
significant impact on the initial entanglement between them \cite{Chowdhury:2021ieg}. However, there 
remain many other extensive areas to venture further in this endeavour. One such area is the 
possibility of entanglement harvesting from the non-vacuum quantum fluctuations, which may be relevant 
from the practical point of view. Till now, so far we are aware of, no investigation related to 
entanglement harvesting with the excited fields has been done. This venture gains its motivation as, 
in nature, it is not guaranteed that the background field will be in a vacuum state. And if the 
background field state is indeed in some excited state, how much do the results concerning the 
entanglement harvesting change compared to the vacuum fluctuations.

In the current work, we are going to investigate the entanglement harvesting condition (see 
\cite{Koga:2018the, Koga:2019fqh, Hu:2015lda} for elaborate discussions on these conditions), for 
two UD detectors in inertial motion interacting with the {\it single-particle background field 
states} (motivated from the work \cite{Lochan:2014xja} where the detector's response function for an 
accelerated detector in single particle excited state has been studied) in $(1+1)$ and $(1+3)$ 
dimensions. In principle the field can be in any excited state; but here, following 
\cite{Lochan:2014xja}, we consider the simplest model where the fields are in single-particle 
excited state. We will see that this becomes an analytically tractable system. In particular, we 
have considered detectors in parallel inertial motion in $(1+1)$ dimensions, and in parallel and 
perpendicular inertial motions in $(1+3)$ dimensions. Our study involves detectors interacting for 
eternity with the background massless, minimally coupled scalar quantum field through monopole-type 
couplings. Here we aim to provide a rigorous formulation for entanglement harvesting with a 
background non-vacuum field state. In this regard, we have followed the works \cite{Koga:2018the, 
Koga:2019fqh}, where one can thoroughly understand entanglement harvesting conditions with detectors 
interacting with the field vacuum. 

In particular, our principal observation in this work is that the entanglement harvested with 
inertial detectors from the single-particle field states is lower than that harvested from the field 
vacuum. In this regard, we consider normalizable single-particle states with two specific types of 
distribution functions. Namely the exponential decaying and the Gaussian distribution functions (see 
\cite{Lochan:2014xja} for elaborate discussions on these field states). We also observe that in both 
 $(1+1)$ and $(1+3)$ dimensions, inertial UD detectors' self transition probabilities are non-zero 
due to the non-vacuum background fluctuations. In $(1+1)$ dimensions, our observations suggest that 
one cannot harvest entanglement between two comoving inertial detectors. In $(1+3)$ dimensions, 
two parallelly moving observers with equal velocity do not harvest entanglement. While in 
perpendicular motion, we do not confirm a similar situation. Furthermore, in this work, we elucidate 
the different characteristics of entanglement harvesting and study its dependence on the parameters 
of our considered system.

This work is organized as follows. In sec. \ref{sec:formulation} we shall provide a brief discussion 
of the model set-up consisting of two UD detectors interacting with a minimally coupled, massless 
scalar field through monopole couplings in the background of a non-vacuum field state. This section 
contains the discussion on the general mathematical description of the entanglement harvesting 
condition. In sec. \ref{sec:Uniform-Velocity-2D} we shall consider two inertial Unruh-DeWitt 
detectors in $(1+1)$ dimensions and investigate the individual detector transition probabilities and 
entanglement harvesting conditions. Subsequently, in sec. \ref{sec:Uniform-Velocity-4D}, UD 
detectors are considered in $(1+3)$ dimensions for studying the same relevant quantities. There are 
two special cases for $(1+3)$ dimensions; one is for parallel motion, and the other is for the 
perpendicular motion of the detectors. They constitute different subsections. We conclude by 
discussing our findings in sec. \ref{sec:discussion}.

\section{The model and working formulas}\label{sec:formulation}

This section presents the formulation for understanding the possibility of two uncorrelated atomic 
Unruh-DeWitt detectors getting entangled over time while interacting with a general field state. Particularly we are interested to find the condition to be entangled and also the quantification of it will be done. The whole analysis will be valid till the second order perturbative series of the density matrix of the system, when the expansion is done order by order in terms of interaction strength.

\subsection{The system: a general framework}
We 
consider two two-level Unruh-DeWitt detectors, associated with our observers Alice and Bob, denoted 
as $A$ and $B$. Furthermore, we perceive the detectors to be point-like and interacting with a 
background massless, minimally coupled real scalar quantum field $\phi(X)$ through monopole interaction. 
Then the interaction action is given by
\begin{eqnarray}
S_{i n t} &=& c \int_{-\infty}^{\infty}\left[ \kappa_{A}\left(\tau_{A}\right) 
m_{A}\left(\tau_{A}\right) \Phi\left(X_{A}\left(\tau_{A}\right)\right) d \tau_{A}\right.\no\\
&& \left.~~~+ \kappa_{B}\left(\tau_{B}\right) m_{B}\left(\tau_{B}\right) 
\Phi\left(X_{B}\left(\tau_{B}\right)\right) d \tau_{B}\right]\,,
\end{eqnarray}
where $c$ is the coupling constant between different detectors and the scalar field (which we have 
assumed to be the same for both the detectors $c_{A}=c_{B}=c$), $\kappa_{j}\left(\tau_{j}\right)$ is 
the switching function for $j^{th}$ detectors (with $j=A,~B$) and $\tau_{j}$ is the individual 
detector's proper time. The initial state of the composite system is taken to be 
$|in\rangle=|\Psi\rangle|E_{0}^{A}\rangle|E_{0}^{B}\rangle$, where $|\Psi\rangle$ is a general field 
state and $|E_{n}^{j}\rangle$ is the $n^{th}$ state of detector $j~(n=0,1)$. In the asymptotic 
future one can obtain the final state of the system as $|out\rangle=T\{e^{iS_{int}}\}|in\rangle$. 
Then by tracing out the field degrees of freedom from the final total density matrix one can get 
the reduced density matrix for the detectors $\rho_{AB}$, which in the basis of 
$\{|E_{1}^{A}\rangle|E_{1}^{B} \rangle,|E_{1}^{A} \rangle|E_{0}^{B}\rangle, 
|E_{0}^{A}\rangle|E_{1}^{B} \rangle,|E_{0}^{A}\rangle|E_{0}^{B}\rangle\}$ is expressed as
\begin{equation}\label{eq:detector-density-matrix}
\rho_{AB}=\begin{pmatrix}
	0			&0					&0				&c^{2}\mathcal{E}\\
	0			&c^{2}\mathcal{P}_{A}		&c^{2}\mathcal{P}_{AB}	&0\\
	0			&c^{2}\mathcal{P}^{\star}_{AB}	&c^{2}\mathcal{P}_{B}	&0\\	
c^{2}\mathcal{E}^{\star}	&0					&0				&1-c^{2}\mathcal{P}_{A}-c^{2}\mathcal{P}_{B}
\end{pmatrix}+O(c^{4}).
\end{equation}
To arrive at this density matrix we have used the expression of the monopole moment 
operator $m_{j}(0) = |E_{0}^{j}\rangle \langle E_{1}^{j}| + |E_{1}^{j}\rangle \langle E_{0}^{j}| $.
The expressions for the matrix elements are
\begin{eqnarray}
\mathcal{P}_{j}&=&\left|\left\langle E_{1}^{j}\left|m_{j}(0)\right| 
E_{0}^{j}\right\rangle\right|^{2} \mathcal{I}_{j}\,, \no\\
\mathcal{E}&=&\left\langle 
E_{1}^{B}\left|m_{B}(0)\right| E_{0}^{B}\right\rangle\left\langle E_{1}^{A}\left|m_{A}(0)\right| 
E_{0}^{A}\right\rangle \mathcal{I}_{\varepsilon}\,,\no \\ 
\mathcal{P}_{A B}&=&\left\langle E_{1}^{A}\left|m_{A}(0)\right| E_{0}^{A}\right\rangle\left\langle 
E_{1}^{B}\left|m_{B}(0)\right| E_{0}^{B}\right\rangle^{\dagger} \mathcal{I}_{A B}\,.
\end{eqnarray}
Again the quantities $\mathcal{I}_{j}$, $\mathcal{I}_{AB}$ and $\mathcal{I}_{\varepsilon}$ in 
these expressions are given by
\begin{eqnarray}\label{IExpressions}
\mathcal{I}_{j}&=&\intinf\intinf{d\tau_{j}d\tau'_{j}}\kappa_j(\tau_j)\kappa_j(\tau'_j)e^{
-i\Delta{E}(\tau_{j}-\tau'_{j})}\no\\&&~~~~~~~~~~~~~~~~~~~~~~~~~~\times		
\langle\Psi|\phi(X_{j})\phi(X'_{j})|\Psi\rangle\,,\no\\ 
\mathcal{I}_{\varepsilon}&=&-\intinf\intinf{d\tau_{B}d\tau'_{A}}\kappa_B(\tau_B)\kappa(\tau'_A)	
e^{i\Delta{E}(\tau'_{A}+\tau_{B})}\no\\&&~~~~~~~~~~~~~~~~~~~~~~~\times	
\langle\Psi|T\phi(X_B)\phi(X'_A)|\Psi\rangle\,,\no\\
\mathcal{I}_{A 
B}&=&\intinf\intinf{d\tau_{B}d\tau'_{A}}\kappa_B(\tau_B)\kappa_A(\tau'_A)e^{i\Delta{E}(\tau'_{A
}-\tau_{B})}\no\\&&~~~~~~~~~~~~~~~~~~~~~~~~~~\times		
\langle\Psi|\phi(X_{B})\phi(X'_{A})|\Psi\rangle\,,
 \end{eqnarray}
where, $\langle\Psi|\phi(X_{i})\phi(X'_{j})|\Psi\rangle$ is the two point function corresponding to 
a general field state.
We should also mention that to arrive at these expressions the explicit forms of the initial field 
state $|\Psi\rangle$ were not needed, which is evident from Eq. (\ref{IExpressions}). Mostly the 
expressions of the time evolution operators were exploited up to this point. In this regard, one may 
note that by replacing $|\Psi\rangle$ by the vacuum state $|0\rangle$, one can get the density 
matrix if the detectors were interacting with the fields in vacuum (or no particle state). 
Nevertheless, one may look into the article \cite{Koga:2018the} for a detailed understanding of the 
procedure to obtain these expressions in an initial field vacuum. The same method has been adopted 
here as well and we landed up a result which is in identical form as obtained for vacuum state; 
except the vacuum state has been replaced by a general excited field state in the correlation 
functions. Therefore the general discussion regarding the entanglement harvesting condition as well 
as measurement of harvested entanglement will exactly follow from previous literature, except the 
explicit measures of the quantities appearing in matrix elements will differ.
The condition for the two detectors to get entangled \cite{Peres:1996dw, Horodecki:1996nc}, is 
obtained when the partial transposition of the reduced density matrix from Eq. 
(\ref{eq:detector-density-matrix}) has negative eigenvalue. This condition is expressed as
\begin{equation}
 \mathcal{P}_{A}\,\mathcal{P}_{B}<|\mathcal{E}|^2~, 
\end{equation}
which can also be cast into the form \cite{Koga:2018the, Koga:2019fqh}
\begin{equation}\label{eq:cond-entanglement}
 \mathcal{I}_{A}\mathcal{I}_{B}<|\mathcal{I}_{\varepsilon}|^2~.
\end{equation}

Once the possibility of entanglement harvesting is arisen (i.e. the condition (\ref{eq:cond-entanglement}) is satisfied), one may 
study different entanglement measures to quantify the harvested entanglement. In this regard, in the 
two qubits case one of the convenient entanglement measures is the concurrence $\mathcal{C}(\rho_{AB})$ 
(see \cite{Koga:2018the, Koga:2019fqh, Hu:2015lda}), which enables one to estimate the entanglement 
of formation $E_{F}(\rho_{AB})$ (see \cite{Bennett:1996gf, Hill:1997pfa, Wootters:1997id, 
Koga:2018the, Koga:2019fqh}). For two-qubits system the concurrence \cite{Koga:2018the} is
\begin{eqnarray}\label{eq:concurrence-gen-exp}
 \mathcal{C}(\rho_{AB}) &=& 
2c^2 \left(|\mathcal{E}|-\sqrt{\mathcal{P}_{A}\mathcal{P}_{B}}\right)+\mathcal{O}(c^4)\nonumber\\
~&=& 2c^2|\langle E_{1}^{B}|m_{B}(0)| E_{0}^{B}\rangle| |\langle 
E_{1}^{A}|m_{A}(0)| E_{0}^{A}\rangle|\nonumber\\
~&& ~~~~~~\times 
\left(|\mathcal{I}_{\varepsilon}|-\sqrt{\mathcal{I}_{A}\mathcal{I}_{B}}
\right)+\mathcal{O} (c^4)~.
\end{eqnarray}
One can notice that the quantities $|\langle E_{1}^{j}|m_{j}(0)| E_{0}^{j}\rangle|$ are completely 
dependent on the detectors' internal structure. The considered background spacetime, scalar fields, 
and the motions of the detectors do not contribute in them. Then from the point of view of 
investigating the entanglement harvesting due to the background spacetime and motions of detectors 
with specific configuration, it seems sufficient to study the nature of 
\begin{equation}\label{eq:concurrence-I}
\mathcal{C}_{\mathcal{I}} = \left(|\mathcal{I} 
_{\varepsilon}| -\sqrt{\mathcal{I}_{A}\mathcal{I}_{B}} \right)
\end{equation} 
in order to understand more about concurrence. We shall indeed be studying this quantity 
$\mathcal{C}_{\mathcal{I}}$ to understand the characteristics of entanglement harvesting in our 
considered system with respect to different parameters.

\subsection{Choice of field state}
We are interested on excited states of field and so in principle $|\Psi\rangle$ can be chosen as any excited field state or combination of all field states. For simplicity of calculation and analysis, 
in this study, we will consider $|\Psi\rangle$ to be consisted of only single exited field states 
corresponding to modes $u_{k}$. Such choice is mainly inspired from \cite{Lochan:2014xja} where the detector's response function for its accelerated motion has been investigated. In $(1+d)$-dimensions, it can be expressed as \cite{Lochan:2014xja}
 \begin{equation}
|\Psi\rangle= \int\frac{d^{d}k}{(2\pi)^{d/2}\sqrt{2\omega_{k}}}f(k)\hat{a}_{k}|0_{u}\rangle~.
 \end{equation}
where $\omega_{k} = |k|$ and $f(k)$ denotes the normalised probability amplitude of the 
distribution, which satisfies the normalization condition
\begin{equation}\label{eq:Normalization-cond-fw}
 \int\frac{d^{d}k}{(2\pi)^{d}{2\omega_{k}}}\left|f(k)\right|^{2}=1~.
 \end{equation}
In a singly excited state the two point function (for explicit derivation, see 
\cite{book:Birrell, Lochan:2014xja}) is given as 
\begin{eqnarray}\label{eq:NVS-Wightman}
\langle\Psi|\Phi(X_{j}) \Phi(X'_{j})| \Psi\rangle &=& G_{W}\left(X_{j},X'_{j}\right)+\Phi_{ {eff 
}}(X_{j}) \Phi_{{eff 
}}^{\star}(X'_{j})\no \\
&&~~~~~~+\Phi_{{eff }}^{\star}(X_{j}) \Phi_{{eff 
}}(X'_{j})\,,
\end{eqnarray}
where $G_{W}\left(X_{j},X'_{j}\right) \equiv \langle 
0_{u}|\Phi\left(X_{j}\right)\Phi\left(X'_{j}\right)|0_{u} \rangle$ denote the positive frequency 
Wightman function with $T_{j}>T_{j'}$. Moreover, $\Phi_{eff}$ is called the effective field, defined 
by \cite{Lochan:2014xja}
\begin{eqnarray}\label{eq:Expression-Phi-eff}
\Phi_{eff}(X_{j})=\int\frac{d^{d}k}{(2\pi)^{d}2\omega_{k}}f(k)e^{ik\cdot X_{j}}.
\end{eqnarray}
Now because of the fact that $f(k)$ is a scalar distribution it is convenient to consider 
$f(k)\equiv f(\omega_{k})$. With this 
line of thought, one may express the time ordered field expectation value as
\begin{eqnarray}\label{eq:NVS-Feynman}
\langle\Psi|T\,\phi(X_{j})\phi(X'_{j})|\Psi\rangle &=& 
\theta(T_{j}-T'_{j})\,\langle\Psi|\phi(X_{j})\phi(X'_{j})|\Psi\rangle \no\\
~&& ~+\, 
\theta(T'_{j}-T_{j})\,\langle\Psi|\phi(X'_{j})\phi(X_{j})|\Psi\rangle\no\\
&=& i\,G_{F}\left(X_{j},X'_{j}\right)+\Phi_{ {eff }}(X_{j}) 
\Phi_{{eff 
}}^{\star}(X'_{j})\no \\
&&~~~~~~+\Phi_{{eff }}^{\star}(X_{j}) \Phi_{{eff 
}}(X'_{j})~,
\end{eqnarray}
where, $\theta(T_{j}-T'_{j})$ denotes the \emph{Heaviside step function}, and we have used the 
expression of 
Eq. (\ref{eq:NVS-Wightman}) to arrive at the last form. One should also note that here 
$G_{F}\left(X_{j},X'_{j}\right) \equiv -i\langle 
0_{u}|T\,\Phi\left(X_{j}\right)\Phi\left(X_{j'}\right)|0_{u} \rangle$ denotes the Feynman 
propagator. 
Thus one can easily observe that the expressions given in 
(\ref{IExpressions}) will contain two contributions -- one from pure vacuum fluctuations and another part is due to the effect of choosing non-vacuum state. Therefore the entanglement harvesting phenomenon in this case effectively driven by vacuum fluctuation of the fields as well as by an effective field configuration emerged due to the non-vacuum property of field state. Hence we expect that the usual entanglement harvesting phenomenon through vacuum fluctuation suffers modification if the background field is in excited state.

Then one may also express the integral $\mathcal{I}_{\varepsilon}$ from Eq. 
(\ref{IExpressions}) as
\begin{equation}\label{eq:Iev-Ienv}
\mathcal{I}_{\varepsilon}=\mathcal{I}_{\varepsilon}^{vac}+\mathcal{I}_{\varepsilon}^{nv}~,
\end{equation}
where
\begin{subequations}
\begin{eqnarray}\label{eq:Iev-Ienv-a}
 \mathcal{I}_{\varepsilon}^{vac} &=& 
-i\intinf\intinf{d\tau_{B}d\tau'_{A}}\,
e^{i\Delta{E}(\tau'_{A}+\tau_{B})}\no\\&&~~~~~~~~~~~~~~~~~~~\times	
G_{F}(X_B,X'_A)~,
\\\label{eq:Iev-Ienv-b}
\mathcal{I}_{\varepsilon}^{nv} &=& -\intinf\intinf{d\tau_{B}d\tau'_{A}}\,
e^{i\Delta{E}(\tau'_{A}+\tau_{B})}\no\\
~&&\times\,[\Phi_{ {eff }}(X_{B}) 
\Phi_{{eff 
}}^{\star}(X'_{A})\no \\
&&~~~~~~+\Phi_{{eff }}^{\star}(X_{B}) \Phi_{{eff 
}}(X'_{A})]~.
\end{eqnarray}
\end{subequations}
Here we have considered trivial switching functions $\kappa_{j}\left(\tau_{j}\right)=1$, which 
signifies that the detectors are eternally interacting with the field \footnote{For 
practical purpose one should choose adiabatic or finite time switching function. But these choice 
normally induces non-analytical evaluation of the integrations and therefore the numerical procedure 
is usually adopted \cite{Henderson:2017yuv, Gallock-Yoshimura:2021yok, Tjoa:2020eqh, Foo:2021gkl, 
Cong:2020nec}. Hence for simplicity and fulfilment of possibility for analytic calculation, here we 
adopted such choice to build our model. Following this spirit the same has been considered earlier as well 
\cite{book:Birrell, Koga:2018the, Koga:2019fqh, Barman:2021bbw, Barman:2021kwg}.}. Furthermore, 
using the relation between Feynman propagator and the Wightman function $iG_{F}\left(X,X'\right) = 
G_{W}\left(X,X'\right) + i G_{R}\left(X',X\right) = G_{W}\left(X,X'\right) + \theta(T'-T) 
\left\{G_{W}\left(X',X\right)-G_{W}\left(X,X'\right)\right\}$ one can simplify the calculation of 
the integral $\mathcal{I}_{\varepsilon}^{vac}$ as
\begin{eqnarray}\label{eq:Ie-integral}
 \mathcal{I}_{\varepsilon}^{vac} &=& \mathcal{I}_{\varepsilon_{W}}^{vac} + 
\mathcal{I}_{\varepsilon_{R}}^{vac},
\end{eqnarray}
with
\begin{subequations}
\begin{eqnarray}\label{eq:Ie-integral-a}
 \mathcal{I}_{\varepsilon_{W}}^{vac} &=& -\int_{-\infty}^{\infty}d\tau_{B} 
\int_{-\infty}^{\infty}d\tau_{A}~\scalebox{0.91}{$e^{i(\Delta 
E^{B}\tau_{B}+\Delta E^{A}\tau_{A})} $}\nonumber\\
~&& ~~~~~~~~~~~~~~~~~\times\,G_{W}(X_{B},X_{A})\,,\\\label{eq:Ie-integral-b}
\mathcal{I}_{\varepsilon_{R}}^{vac} &=& -\int_{-\infty}^{\infty}d\tau_{B} 
\int_{-\infty}^{\infty}d\tau_{A}~\scalebox{0.91}{$e^{i(\Delta 
E^{B}\tau_{B}+\Delta E^{A}\tau_{A})} $} \theta(T_{A}-T_{B})\no\\
~&& ~~~\times\,\big\{G_{W}\left(X_{A},X_{B}\right)-G_{W}\left(X_{B},X_{A}\right)\big\}\,,
\end{eqnarray}
\end{subequations}
where, $G_{R}\left(X,X'\right) \equiv i\theta(T-T')\langle 
0_{u}|\left[\Phi\left(X'\right),\Phi\left(X\right)\right]|0_{u}
\rangle$ signifies the retarded Green's function.
We will use the above form for our purpose to evaluate $\mathcal{I}_{\varepsilon}$. It is observed 
that one only needs the expressions of the integrals $\mathcal{I}_{A}$, $\mathcal{I}_{B}$  and 
$\mathcal{I}_{\varepsilon}$ for verification of the condition (\ref{eq:cond-entanglement}) for 
entanglement harvesting. Furthermore, we also note that with the help of Eq. (\ref{eq:NVS-Wightman}) 
one can express the integral $\mathcal{I}_{j}$ from (\ref{IExpressions}) as
\begin{equation}\label{eq:Ijv-Ijnv}
\mathcal{I}_{j}=\mathcal{I}_{j}^{vac}+\mathcal{I}_{j}^{nv}~,
\end{equation}
where $\mathcal{I}_{j}^{vac}$ exactly denotes the contribution if the background field states were 
vacuum rather than non-vacuum. Whereas, $\mathcal{I}_{j}^{nv}$ denotes the extra contribution arriving 
due to the consideration of the non-vacuum background field states. One may express these 
quantities as
\begin{subequations}
\begin{eqnarray}\label{eq:Ijv-Ijnv-a}
 \mathcal{I}_{j}^{vac} &=& \intinf\intinf d\tau_{j}\,d\tau'_{j}\,e^{
-i\Delta{E}(\tau_{j}-\tau'_{j})}\,\no\\
~&& ~~~~~~~~~~~~~\times~G_{W}(X_{j},X'_{j})~,
\\\label{eq:Ijv-Ijnv-b}
\mathcal{I}_{j}^{nv} &=& |\mathcal{A}(\Delta E)|^2 + |\mathcal{B}(\Delta E)|^2~,
\end{eqnarray}
\end{subequations}
where, 
\begin{subequations}
\begin{eqnarray}\label{eq:Ijv-Ijnv-2a}
\mathcal{A}(\Delta E) &=& \intinf d\tau'_{j}\,e^{
i\Delta{E}\,\tau'_{j}}\,\Phi_{eff}(\tau'_{j})~, \\\label{eq:Ijv-Ijnv-2b}
~\mathcal{B}(\Delta E) &=&  \intinf d\tau'_{j}\,e^{
i\Delta{E}\,\tau'_{j}}\,\Phi_{eff}^{\star}(\tau'_{j})~.
\end{eqnarray}
\end{subequations}
We shall use these expressions for our purpose to obtain the explicit expression 
of $\mathcal{I}_{j}$.

\section{Entanglement harvesting: Motion in $(1+1)$ dimensions}\label{sec:Uniform-Velocity-2D}

Here we consider two atomic Unruh-DeWitt detectors in uniform velocity in $(1+1)$ dimensional 
spacetime. We assume Alice and Bob to have velocities $v_{A}$ and $v_{B}$ respectively. In this scenario one is able to 
express the Minkowski time $t_j$ and position $x_j$ (i.e., we are now considering 
$X_{j}=(t_{j},x_{j})$) of these two detectors in terms of their respective proper times $\tau_{j}$ 
as
\begin{eqnarray}\label{eq:Lorentz-trans-1p1}
 t_{j} &=& \gamma_{j}\,\tau_{j}~,\nonumber\\
 ~x_{j} &=& v_{j}\,\gamma_{j}\,\tau_{j}~,
\end{eqnarray}
where, $j$ denotes either $A$ (corresponds to Alice) or $B$ (corresponds to Bob), and $\gamma_{j}$ 
is the Lorentz factor, $\gamma_{j}=1/\sqrt{1-v_{j}^2}$.
\begin{figure}[!h]
\centering
\includegraphics[width=0.42\textwidth]{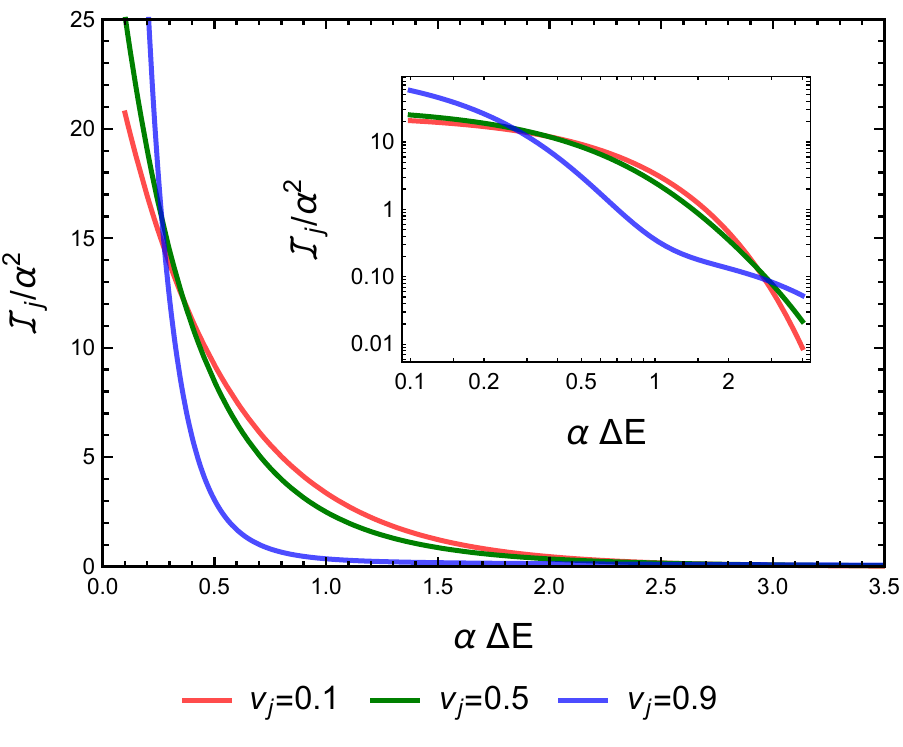}
\hskip 10pt
\includegraphics[width=0.42\textwidth]{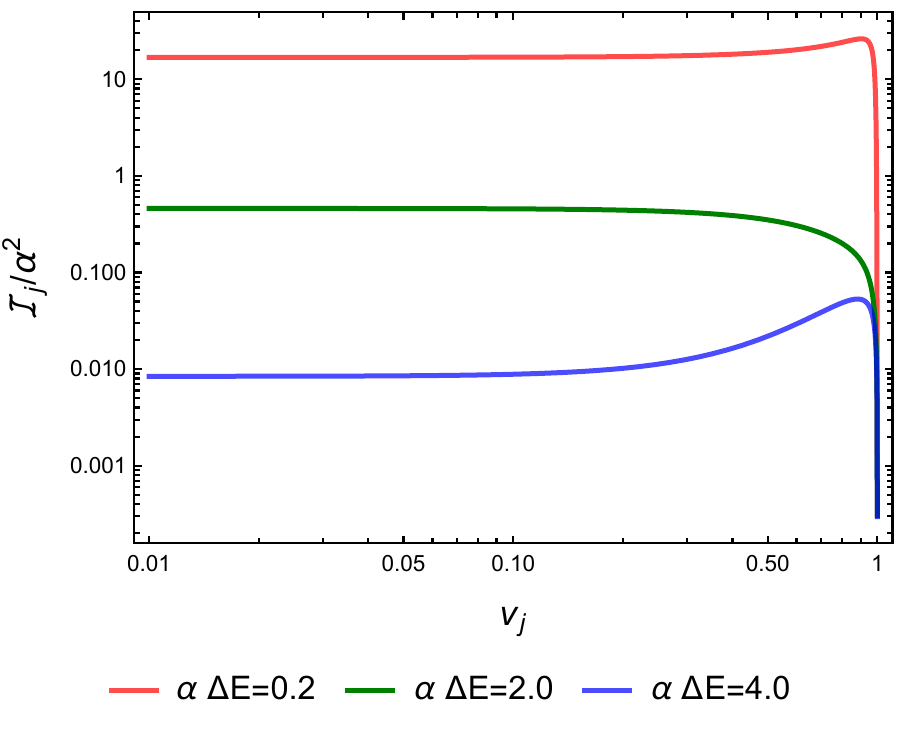}
\caption{In $(1+1)$ dimensions the integral $\mathcal{I}_{j}$ denoting individual 
detector transition probability is plotted as functions of $\alpha\,\Delta E$ and $v_{j}$ in the 
upper and the lower figures respectively. For both the cases we considered the distribution function 
$f(\omega_{k})=C\, \omega_{k}\,e^{-\alpha\omega_{k}}$. From the upper plot one can observe that in 
the low $(\alpha\,\Delta E\lesssim 0.26)$ and high $(\alpha\,\Delta E\gtrsim 2.9)$ value regions of 
$\alpha\,\Delta E$ the transition probability increases with increasing detector velocity upto certain large velocities. While in 
the intermediate region it decreases with increasing velocity.  We mention that the upper inner and the lower plots are presented in \emph{Log-Log} fashion.}
\label{fig:Ia-1p1-ED}
\end{figure}
\begin{figure}[!h]
\centering
\includegraphics[width=0.42\textwidth]{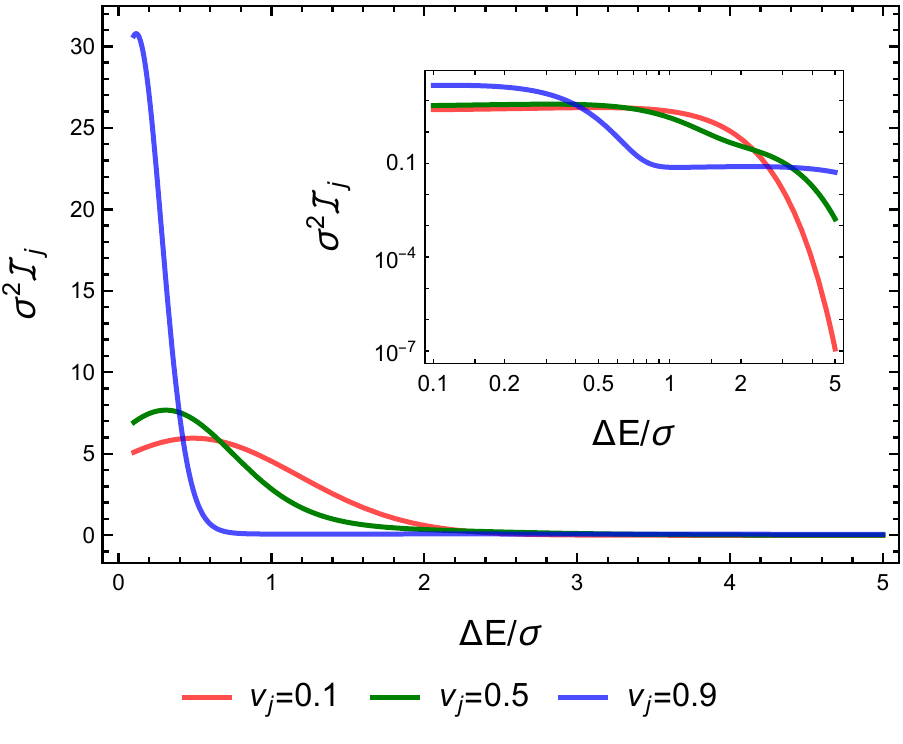}
\hskip 10pt
\includegraphics[width=0.42\textwidth]{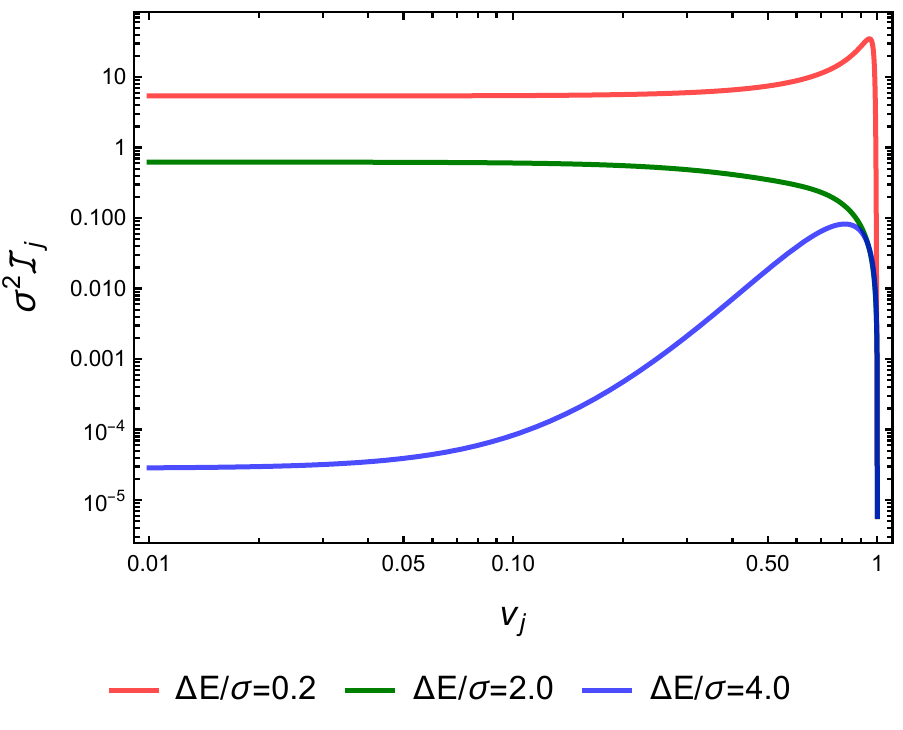}
\caption{In $(1+1)$ dimensions the integral $\mathcal{I}_{j}$ denoting individual 
detector transition probability is plotted as functions of $\Delta E/\sigma$ and $v_{j}$ in the 
upper and the lower figures respectively. For both the cases we considered the distribution function 
$f(\omega_{k})=C\, \omega_{k} ~e^{-(\omega_{k}-\omega_{0})^{2}/2\sigma^{2}}$, with 
$\omega_{0}/\sigma=0.5$. From the upper plot one can observe that in the low $(\Delta 
E/\sigma\lesssim 0.38)$ and high $(\Delta E/\sigma\gtrsim 3.3)$ value regions of $\Delta E/\sigma$ 
the transition probability increases with increasing detector velocity upto certain velocities. While in the intermediate 
region it decreases with increasing velocity. Here 
also we have presented the upper inner and the lower plots in \emph{Log-Log} fashion.}
\label{fig:Ia-1p1-GF}
\end{figure}
\begin{figure}[!h]
\centering
  \includegraphics[width=0.42\textwidth]{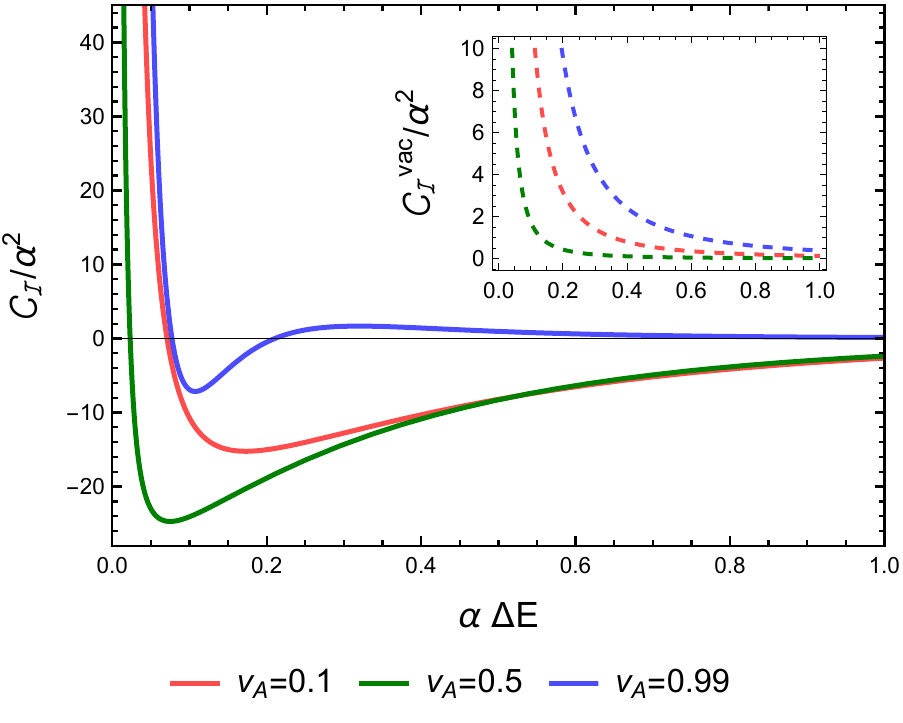}
  \hskip 10pt
  \includegraphics[width=0.42\textwidth]{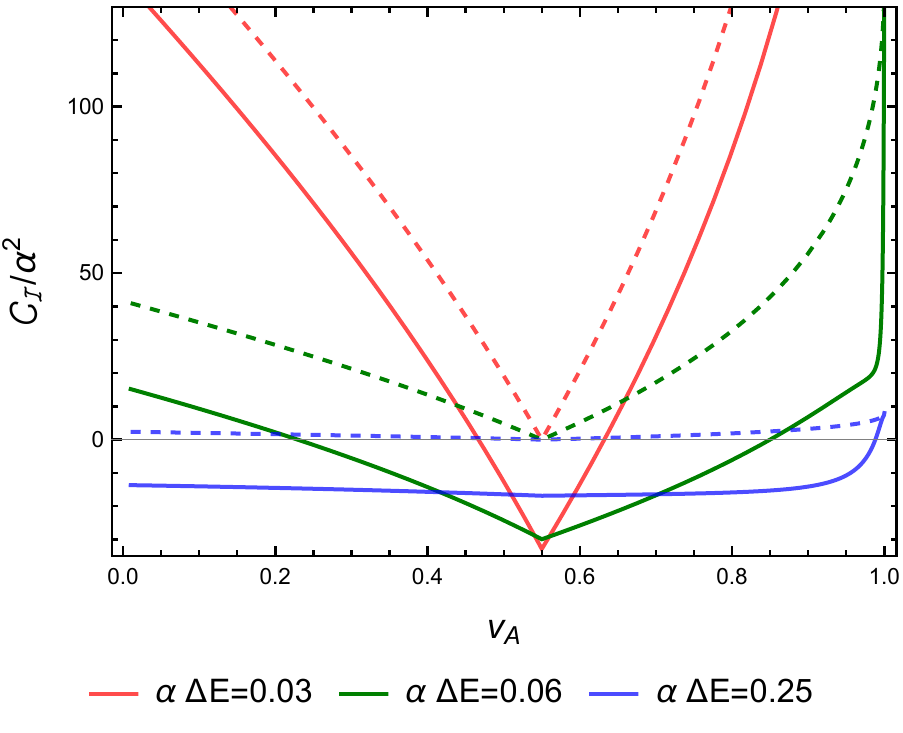}
\caption{In $(1+1)$ dimensions the quantity $\mathcal{C}_{\mathcal{I}}$ signifying 
the concurrence (the solid lines) is plotted as functions of $\alpha\,\Delta E$ and $v_{A}$ in the 
upper and the lower plots respectively. Here the velocity of detector $B$ is fixed at $v_{B}=0.55$.
To arrive at these plots we have considered the 
distribution function $f(\omega_{k})=C\, \omega_{k}\,e^{-\alpha\omega_{k}}$. We also mention that 
$\mathcal{C}_{\mathcal{I}}^{vac}$, denoted by the dashed lines, indicates the concurrence if the 
detectors were interacting with the field vacuum. In the lower plot we have chosen the fixed parameter values for $\alpha\,\Delta E$ conveniently as suggested from the upper plot.}
\label{fig:C-1p1-ED}
\end{figure}
\begin{figure}[!h]
\centering
\includegraphics[width=0.42\textwidth]{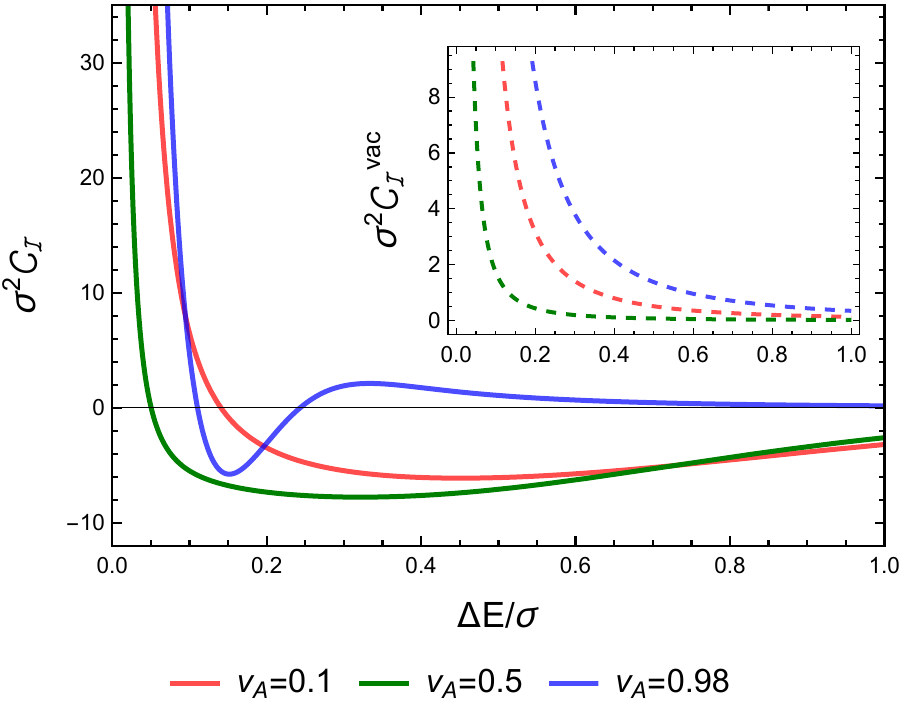}
\hskip 10pt
\includegraphics[width=0.42\textwidth]{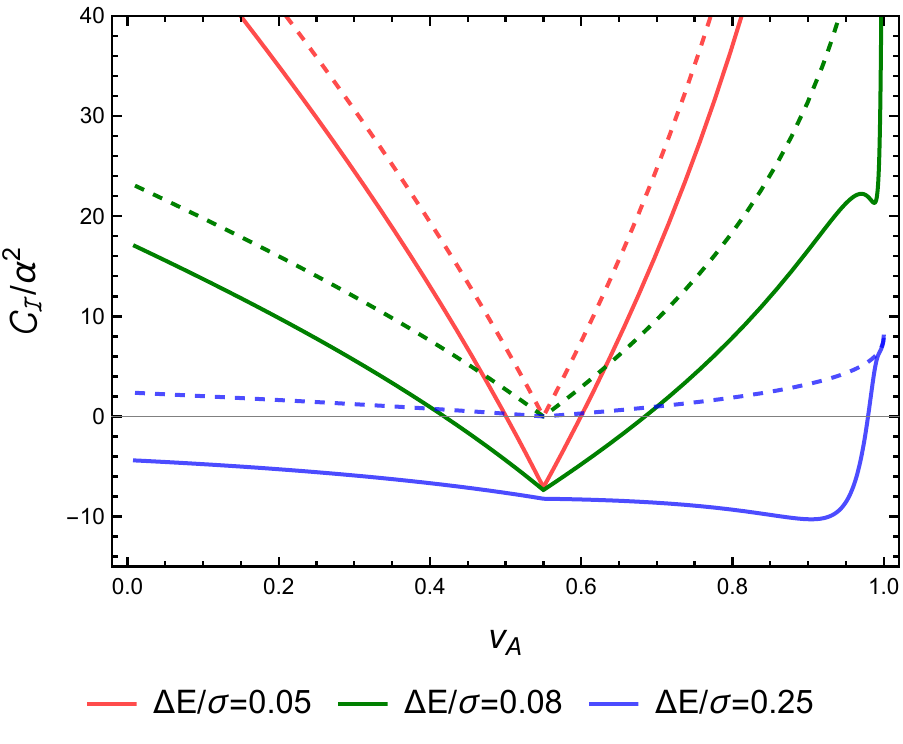}
\caption{In $(1+1)$ dimensions the quantity $\mathcal{C}_{\mathcal{I}}$ signifying 
the concurrence (the solid lines) is plotted as functions of $\Delta E/\sigma$ and $v_{A}$ in the 
upper and the lower plots respectively. We have considered the distribution function 
$f(\omega_{k})=C\, \omega_{k} ~e^{-(\omega_{k}-\omega_{0})^{2}/2\sigma^{2}}$ in this case with 
$\omega_{0}/\sigma=0.5$ and $v_{B}=0.55$. Here also $\mathcal{C}_{\mathcal{I}}^{vac}$ (the dashed lines) indicates 
the concurrence if the detectors were interacting with the field vacuum.}
\label{fig:C-1p1-GF}
\end{figure}

\subsection{Evaluation of $\mathcal{I}_{j}$}
The integrals $\mathcal{I}_{j}$ from the quantities $\mathcal{P}_{j}$ signify individual detector 
transition probabilities. We shall be first evaluating these integrals. Note that in $(1+1)$ 
dimensions the expression of the scalar field mode function is given by $u_{k}(X_{j}) = 
(1/\sqrt{4\pi\omega_{k}})\,e^{-i\,\omega_{k}\,t_{j}+i\,k\,x_{j}}$ and so the Wightman function is
\begin{eqnarray}
 G_{W}(X_{j},X'_{j}) = 
\intinf\frac{dk}{4\pi\omega_{k}}\,e^{-i\,\omega_{k}\,(t_{j}-t'_{j})+i\,k\,(x_{j}-x'_{j})}\,.
\end{eqnarray}
Utilizing this Wightman function with the coordinate transformation from Eq. 
(\ref{eq:Lorentz-trans-1p1}) one can find out the quantity $\mathcal{I}_{j}^{vac}$ from 
(\ref{eq:Ijv-Ijnv-a}) as
\begin{eqnarray}\label{eq:IjVac-1p1}
 \mathcal{I}_{j}^{vac} &=& \intinf\frac{dk}{4\pi\omega_{k}}\,\intinf\intinf 
d\tau_{j}\,d\tau'_{j}\,e^{
-i\Delta{E}(\tau_{j}-\tau'_{j})}\no\\
~&&~~~~~~~~~~~~~~\times\,e^{-i\,(\omega_{k}-k\,v_{j})\,\gamma_{j}(\tau_{j}-\tau'_{j})}\,\no\\
~&=& \intinf\,\frac{\pi\,dk}{\omega_{k}}\, \left\{\delta[\gamma_{j}\,(\omega_{k}-k\,v_{j})+\Delta 
E]\right\}^2~\no\\
~&=& \int_{0}^{\infty}\,\frac{\pi\,d\omega_{k}}{\omega_{k}}\,\Big[ 
\left\{\delta[\gamma_{j}\,\omega_{k}(1-\,v_{j})+\Delta 
E]\right\}^2~\no\\
&& ~~~~~~~~+\, 
\left\{\delta[\gamma_{j}\,\omega_{k}(1+\,v_{j})+\Delta 
E]\right\}^2\Big]\,,
\end{eqnarray}
where $\omega_{k} = |k|$ has been used. The final expression in Eq. (\ref{eq:IjVac-1p1}) vanishes as 
we have $0\le v_{j}\le1$ and $\Delta E>0$. 
The reason behind is, in that situation the argument of the Dirac delta distribution is always 
positive in the considered integration range of $\omega_k$. This is expected as an inertial detector does not suffer transition due to vacuum fluctuation of fields.

On the other hand, if one uses the expression 
of $\Phi_{eff}(X)$ from Eq. (\ref{eq:Expression-Phi-eff}) for $(1+1)$ dimensions, then the quantity 
$\mathcal{A}(\Delta E)$, given by Eq. (\ref{eq:Ijv-Ijnv-2a}), becomes
\begin{eqnarray}\label{eq:ADE-1p1-A&B}
\mathcal{A}(\Delta E) &=&\intinf\frac{f(\omega_{k})\,dk}{4\pi\omega_{k}}\intinf d\tau'_{j}\,e^{
i[\Delta{E}\,-\,(\omega_{k}-k\,v_{j})\,\gamma_{j}]\tau'_{j}}\no\\
&=&\intinf\frac{dk}{2\omega_{k}}\,f(\omega_{k})\,\delta[\gamma_{j}\,(\omega_{k}-k\,v_{j})-\Delta{E}]
\no\\
~&=& \frac{1}{2\,\Delta E} \,\left\{f(D_{j}\,\Delta E) + f(\Delta 
E/D_{j})\right\}~,
\end{eqnarray}
where, $D_{j} = \sqrt{1+v_{j}}/\sqrt{1-v_{j}}$. In a similar manner one can find out the other 
quantity from (\ref{eq:Ijv-Ijnv-2b}) as
\begin{eqnarray}
 \mathcal{B}(\Delta E) &=& \intinf\frac{f(\omega_{k})\,dk}{4\pi\omega_{k}}\intinf d\tau'_{j}\,e^{
i[\Delta{E}\,+\,(\omega_{k}-k\,v_{j})\,\gamma_{j}]\tau'_{j}}\no\\
&=&\intinf\frac{dk}{2\omega_{k}}\,f(\omega_{k})\,\delta[(\omega_{k}-k\,v_{j})\,\gamma_{j}+\Delta{E}]
\no\\
~&=& 0~,
\label{BRM1}
\end{eqnarray}
for $\Delta E>0$. Therefore, in $(1+1)$ dimensions with two detectors in uniform velocities the 
quantity $\mathcal{I}_{j}$ from Eq. (\ref{eq:Ijv-Ijnv}) is entirely given by $\mathcal{A}(\Delta 
E)$, specifically as $\mathcal{I}_{j}(\Delta E) = |\mathcal{A}(\Delta E)|^2$. Once the explicit form 
of distribution function $f(\omega_k)$ is given, then we will be able to know 
$\mathcal{I}_{j}(\Delta E)$. One may look into Appendix \ref{Appn:dist-fns-1p1} for the expressions 
of a few possible distribution functions $f(\omega_{k})$ in $(1+1)$ dimensions. We choose them to be 
similar that have been adopted in \cite{Lochan:2014xja} to investigate the single detector's 
response function in an accelerated frame.

Note that the quantities $\mathcal{I}_{j}$ denote the single detector transition probabilities. 
Since we have not found any discussions in the literature on transition probabilities of single 
inertial detectors in flat spacetime interacting with the non-vacuum field states, here we end this 
subsection with an analysis on the features of this quantity. In particular, in Fig. 
\ref{fig:Ia-1p1-ED} and \ref{fig:Ia-1p1-GF} we have plotted these quantities for the specific 
distribution functions $f(\omega_{k})=C\, \omega_{k}\,e^{-\alpha\omega_{k}}$ and $f(\omega_{k})=C\, 
\omega_{k} ~e^{-(\omega_{k}-\omega_{0})^{2}/2\sigma^{2}}$ respectively. 
Here $C$ is the normalization constant and its explicit form, which is imperative 
for obtaining the numerical value of $\mathcal{I}_{j}$, is determined in Appendix 
\ref{Appn:dist-fns-1p1}.
%
%

From these figures, with both the exponentially damping and Gaussian distribution functions, one can 
observe that in different regions of $\Delta E$ the transition probability varies differently with 
the velocity $v_{j}$ of the detector. For example, the upper plots of Fig. \ref{fig:Ia-1p1-ED} and 
\ref{fig:Ia-1p1-GF} signify that for very low and high $\Delta E$ the transition increases with 
increasing $v_j$. While in the intermediate range of $\Delta E$, transition decreases with 
increasing $v_j$.
The behavior of the transition probability with respect to the velocity $v_{j}$ in different $\Delta 
E$ regimes is further elucidated in the lower plots of the concerned figures. From those plots, one 
further can observe that in the low and high $\Delta E$ regime, when $\Delta E/\kappa = 0.2$ and 
$\Delta E/\kappa = 4$ (here $\kappa$ can be $\sigma$ or $1/\alpha$ depending on the considered 
distribution function), the transition rate initially increases with increasing velocity $v_{j}$ up 
to certain values of $v_j$ and then decreases. While this characteristic is different in the 
intermediate transition energy regimes, e.g., when $\Delta E/\kappa = 2$. In the later case 
transition probability decreases with the increase of $v_j$.

Let us now elucidate on the reasoning behind the specific features of the lower plots in Figs. \ref{fig:Ia-1p1-ED} and \ref{fig:Ia-1p1-GF}, which is due to the appearance of two terms in Eq. (\ref{eq:ADE-1p1-A&B}). One 
is providing red shift in $\Delta E$ while the other one induces blue shift. In order to understand 
influence of these two effects in the transition amplitude, it is convenient to pay attention on a 
particular distribution of $f(\omega_k)$. Here we consider the exponentially damping one, where
$\mathcal{A}(\Delta E)$ is given by
\begin{eqnarray}
\mathcal{A}(\Delta E) &=& \frac{C}{2} \bigg[\sqrt{\frac{1+v_j}{1-v_j}} e^{-\alpha \Delta E 
\sqrt{\frac{1+v_j}{1-v_j}}}
\nonumber
\\
&+& \sqrt{\frac{1-v_j}{1+v_j}} e^{-\alpha \Delta E \sqrt{\frac{1-v_j}{1+v_j}}}\,\bigg]~.
\label{BRM1}
\end{eqnarray}
Since we have $v_{j}<1$, the above can be expanded 
in Taylor series. Keeping terms up to leading order in $v_j$ (which is here $v_j^2$) one finds
\begin{eqnarray}
\mathcal{A}(\Delta E) \simeq C e^{-\alpha\Delta E}\Big[1+\left(\alpha^{2}\Delta E^2-3\alpha\Delta E+1\right) v_{j}^2\Big]\,.
\label{BRM2}
\end{eqnarray}
Therefore when $\left(\alpha^{2}\Delta E^2-3\alpha\Delta E+1\right)>0$, the magnitude of the 
transition amplitude will increase with the increase of $v_j$. But if $\left(\alpha^{2}\Delta E^2-3\alpha\Delta E+1\right)<0$, then second term in the above expression provides diminishing effect in amplitude. As a 
result the transition amplitude decreases with the increase in $v_j$. This discussion provides 
explanation for the nature of curves for $\alpha\Delta E = 0.2, 2.0$ and $4.0$. But when $v_j$ is large, i.e., $v_{j}\to 1$, this approximation is not valid. In that case it  is noted that the first term in 
(\ref{BRM1}) decays exponentially with the increase of $v_j$. Because of the pre-factor in the 
second exponential, the whole term also decreases. Hence effectively the amplitude decreases with 
the increase of $v_j$. Furthermore, one can check by taking the limit  $v_{j}\to 1$ that the transition amplitude $\mathcal{A}(\Delta E)$ vanishes. Similar things are also happening for the Gaussian distribution  in the lower plot of Fig. \ref{fig:Ia-1p1-GF}.

\subsection{Evaluation of $\mathcal{I}_{\varepsilon}$, and the concurrence 
$\mathcal{C}_\mathcal{I}$}

Let us now proceed to evaluate the integral $\mathcal{I}_{\varepsilon}$. We first consider 
the expression of $\mathcal{I}_{\varepsilon}$ from Eq. (\ref{eq:Iev-Ienv}) and evaluate 
$\mathcal{I}_{\varepsilon}^{vac}$, which again is expressed as (\ref{eq:Ie-integral}). In Eq. 
(\ref{eq:Ie-integral}) the term containing only the Wightman function can be expressed as
\begin{eqnarray}\label{eq:IeWvac-1p1}
 \mathcal{I}_{\varepsilon_{W}}^{vac} 
&=& -\frac{1}{\gamma_{A}\gamma_{B}} \intinf 
\frac{dk}{4\pi\omega_{k}}\,\int_{-\infty}^{\infty}dt_{B} 
\int_{-\infty}^{\infty}dt_{A}~\no\\
~&& \times~ e^{
-i(\omega_{k}-k\,v_{B})\,t_{B}
+i(\omega_{k}-k\,v_{A})\,t_{A}}\no\\
~&& ~~~~~~~~~~~\times\,e^{i\Delta E(t_{B}/\gamma_{B}+t_{A}/\gamma_{A})} \,\no\\
&=& -\frac{\pi}{\gamma_{A}\gamma_{B}} \intinf 
\frac{dk}{\omega_{k}}\,\delta[(\omega_{k}-k\,v_{A})+\Delta{E}/\gamma_{A}]\no\\
~&& \times~ \,\delta[(\omega_{k}-k\,v_{B})-\Delta{E}/\gamma_{B}] \,,
\end{eqnarray}
which will vanish due to the first Dirac-delta distribution for $\Delta{E}>0$. We also mention that 
here we have used the relation $t_{j} = \gamma_{j}\,\tau_{j}$ from Eq. (\ref{eq:Lorentz-trans-1p1}). 
On the other hand, from Eq. (\ref{eq:Ie-integral}) the term containing the retarded Green's function 
can be expressed as
\begin{eqnarray}\label{eq:IeRvac-1p1-1}
&&  \mathcal{I}_{\varepsilon_{R}}^{vac} 
\no\\
&&= \frac{i}{\gamma_{A}\gamma_{B}}\intinf 
\frac{dk}{2\omega_{k}} 
\,\Bigg[\frac{\delta[k(v_{A}-v_{B})-\Delta{E}(1/\gamma_{A}+1/\gamma_{B})]}{\omega_{k}-k 
\,v_{B}-\Delta{E}/\gamma_{B}} \no\\
~&& ~~~+~ \frac{\delta[k(v_{A}-v_{B})+\Delta{E}(1/\gamma_{A}+1/\gamma_{B})]}{\omega_{k}-k 
\,v_{B}+\Delta{E}/\gamma_{B}}\Bigg] \,.
\end{eqnarray}
See Appendix \ref{Appn:1p1-integrals} for an explicit derivation of this expression. 
Here we have also used the fact that the Minkowski time $t_{j}$ are the temporal coordinates 
$T_{j}$ for field mode expansion in (\ref{eq:Ie-integral-b}). For $v_{A}>v_{B}$ this expression 
leads to 
\begin{eqnarray}\label{eq:IeRvac-1p1-2}
\mathcal{I}_{\varepsilon_{R}}^{vac} &=& \frac{i\gamma_{A}\gamma_{B}(v_{A}-v_{B})D_{A}D_{B}}{\Delta 
E^2(D_{A}+D_{B})^2}\,~.
\end{eqnarray}
On the other hand, for $v_{B}>v_{A}$ the integral from Eq. (\ref{eq:IeRvac-1p1-1}) becomes
\begin{eqnarray}\label{eq:IeRvac-1p1-3}
\mathcal{I}_{\varepsilon_{R}}^{vac} &=& \frac{i\gamma_{A}\gamma_{B}(v_{B}-v_{A})D_{A}D_{B}}{\Delta 
E^2(D_{A}+D_{B})^2}~,
\end{eqnarray}
which is basically the same compared to the expression from $v_{A}>v_{B}$ case with 
a change in sign. From the previous sub-section we have seen that the integral of the form $\intinf 
d\tau\,e^{i\Delta{E}\,\tau}\,\Phi_{eff}^{\star}(\tau)$ vanishes for detectors in uniform velocity in 
$(1+1)$ dimensions (see Eq. (\ref{BRM1})). Then, one can perceive that the quantity 
$\mathcal{I}_{\varepsilon}^{nv}$ from Eq. (\ref{eq:Iev-Ienv-b}) will also vanish in this scenario. 
Therefore, in this case we have $\mathcal{I}_{\varepsilon}$ entirely given by 
$\mathcal{I}_{\varepsilon_{R}}^{vac}$, i.e., $\mathcal{I}_{\varepsilon} = 
\mathcal{I}_{\varepsilon_{R}}^{vac}$. Moreover, since there is only a sign change between the 
expressions from Eq. (\ref{eq:IeRvac-1p1-2}) and (\ref{eq:IeRvac-1p1-3}), in both the cases 
$v_{A}>v_{B}$ and $v_{A}<v_{B}$ the quantity $|\mathcal{I}_{\varepsilon}|$ appearing in 
(\ref{eq:concurrence-I}) yields the same feature. Also observe from Eq. (\ref{eq:IeRvac-1p1-2}) and 
(\ref{eq:IeRvac-1p1-3}) that when $v_{A}=v_{B}$ the integral $\mathcal{I}_{\varepsilon}$ vanishes 
and so vanishes the concurrence. Thus in a situation when the two UD detectors are co-moving in 
$(1+1)$ dimensions there will be no entanglement harvesting.
On the other hand, when the velocity of detector $A$ approaches the velocity of 
light, i.e., when $v_{A}\to 1$, the quantity $|\mathcal{I}_{\varepsilon_{R}}^{vac}|= 
1/(2\Delta E^2)$. Therefore, in this limit there is an upper bound in 
the value of the concurrence depending on the value of $v_{B}$ and $\Delta E$. This expressions 
also suggests that in this limit the amount of harvested concurrence (see Eq. 
(\ref{eq:concurrence-I})) should be increasing with decreasing $\Delta E$.

In Fig. \ref{fig:C-1p1-ED} and \ref{fig:C-1p1-GF} we have plotted the concurrence denoting quantity 
$\mathcal{C}_{\mathcal{I}}$ as described in Eq. (\ref{eq:concurrence-I}) considering the 
exponentially damping and Gaussian distribution functions respectively. For a discussion on these 
distributions in $(1+1)$ dimensions see Appendix \ref{Appn:dist-fns-1p1}. We also mention that if 
the detectors were to interact with the field vacuum rather than the singly excited field state, the 
quantities $\mathcal{I}_{j}$ denoting single detectors' transition probabilities would have been 
zero. In that scenario the concurrence is entirely given by $\mathcal{C}_{\mathcal{I}}^{vac} = 
|\mathcal{I}_{\varepsilon}| = |\mathcal{I}_{\varepsilon_{R}}^{vac}|$, a depiction of which is also 
included in these plots. From these figures, one notices that when $v_{A} = v_{B}$, the entanglement 
harvesting, like vacuum field state, ceases to exist for non-vacuum field state as well. Fig. 
\ref{fig:C-1p1-ED} and \ref{fig:C-1p1-GF} also confirms the possibility of entanglement harvesting 
between inertial UD detectors even from the single particle field states in $(1+1)$ dimensions. 
However, comparing the plots for $\mathcal{C}_{\mathcal{I}}$ and $\mathcal{C}_{\mathcal{I}}^{vac}$ 
from these figures, one can clearly see that entanglement harvesting is reduced in the non-vacuum 
case.
In the non-vacuum situation the reduction happens not only in magnitude of concurrence, also the 
range of velocity decreases. A common feature observed with both the distributions (the exponential 
damping and Gaussian distributions) is that the regime of low detector 
transition energy corresponds to quantitatively higher entanglement extraction. 
One should also note that for large fixed transitions energies there is no entanglement harvesting from the low velocity regimes. However, the amount of the harvested entanglement in the high velocity regime keeps decreasing with increasing $\Delta E$.
This is because in 
the very large $v_A$ limit the entangling term varies as $\sim 1/(\Delta E)^2$ and hence 
$\mathcal{I}_\varepsilon$ increase more than $\mathcal{I}_j$ with the decrease of $\Delta E$.

\section{Entanglement harvesting: motion in $(1+3)$ dimensions}\label{sec:Uniform-Velocity-4D}

Now we are going to investigate the same in $(1+3)$ dimensions. Since there are three spacial 
directions, the observers can move in more than one possible directions. We mainly concentrate on 
following motions of the two detectors -- both are moving along the same direction and one is moving 
in the perpendicular direction of the other's motion.

\subsection{Parallel motion}\label{subsec:1p3D-parallel-UV}

In this part we we are going to consider Unruh-DeWitt detectors in parallel uniform velocities in 
$(1+3)$ dimensions. For simplicity we shall consider the detectors to be in motion along the $z$ 
direction. In particular, the observer with detector $A$ is assumed to have an 
inertial motion such that its coordinates are related to its proper time as
\begin{equation}\label{eq:ParaTrajecA}
t_{A}=\gamma_{A} \tau_{A};~~x_{A}=0;~~y_{A}=0;~~z_{A}=\gamma_{A} v_{A}\tau_{A}.
\end{equation}
On the other hand, the trajectory of detector $B$ is
\begin{equation}\label{eq:ParaTrajecB}
t_{B}=\gamma_{B} \tau_{B};~~x_{B}=x_{0};~~y_{B}=y_{0};~~z_{B}=\gamma_{B} v_{B}\tau_{B},
\end{equation}
such that $r_{0}=\sqrt{x_{0}^2+y_{0}^2}$ is the perpendicular distance between the two detectors 
$A$ and $B$. In $(1+3)$ dimensions the effective field $\Phi_{\text {eff }}$ with respect to the 
Minkowski modes is given as
\begin{equation}
\Phi_{\mathrm{eff}}(x)=\int \frac{d^{3}k}{(2 \pi)^{3}} \frac{f(\omega_{k})}{2 \omega_{k}} e^{i k 
\cdot x}~,
\end{equation}
where, $k\cdot x \equiv k_{a}x^{a} = -\omega_{k}t+\vec{k}\cdot\vec{x}~$. We shall see that unlike 
the previous $(1+1)$ dimensional scenario, here the expressions of $\mathcal{I}_{A}$ and 
$\mathcal{I}_{B}$ are not the same due to a finite non-zero initial separation between the two 
detectors.

\begin{figure}[!h]
\centering
\includegraphics[width=0.42\textwidth]{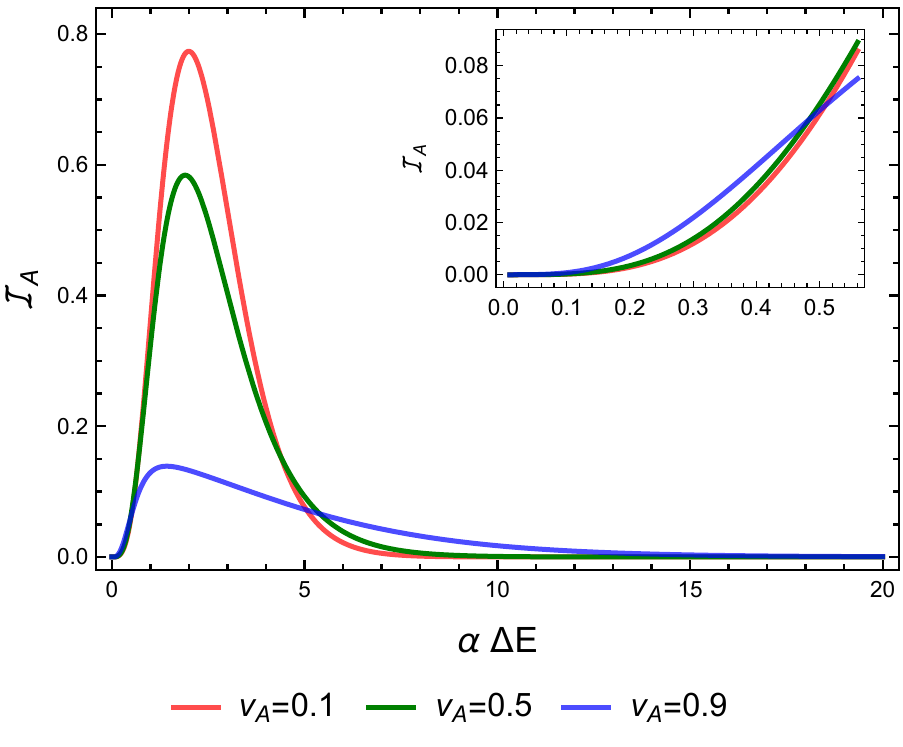}
\hskip 10pt
\includegraphics[width=0.42\textwidth]{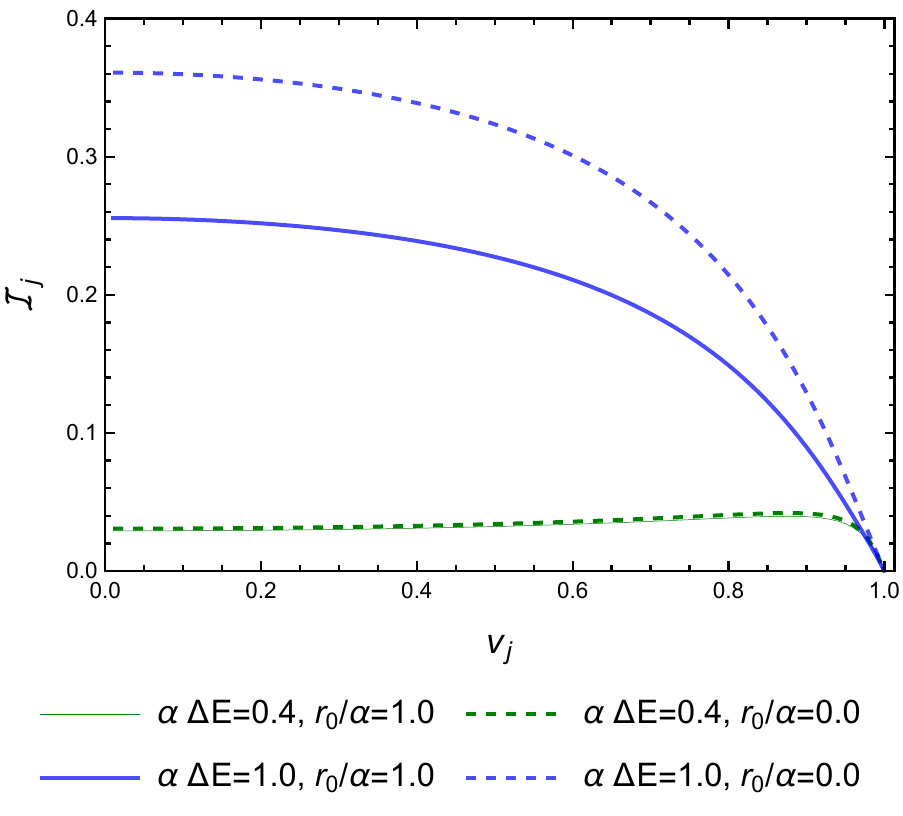}
\caption{In the upper figure the integral $\mathcal{I}_{A}$ (which signifies 
individual detector transition probability) is plotted as a function of $\alpha\,\Delta E$ for fixed 
values of $v_{A}$ in $(1+3)$ dimensions. On the other hand, in the lower figure $\mathcal{I}_{j}$ is plotted with respect to $v_{j}$ for different fixed values of 
$\alpha\,\Delta E$ and $r_{0}/\alpha$. In both of these figures we have considered the distribution 
function $f(\omega_{k})=C~\omega_{k}~e^{-\alpha\omega_{k}}$. It is to be noted that in the lower 
plot $r_{0}/\alpha=1$ corresponds to detector $B$ and $r_{0}/\alpha=0$ corresponds to detector $A$. 
From the upper figure one can observe that in different regions of $\alpha\,\Delta E$ the transition 
probability may increase or decrease with increasing detector velocity. This observation is similar to 
the ones from $(1+1)$ dimensions, see Fig. \ref{fig:Ia-1p1-ED}.}
\label{fig:Ij-1p3-ED}
\end{figure}
\begin{figure}[!h]
\centering
\includegraphics[width=0.42\textwidth]{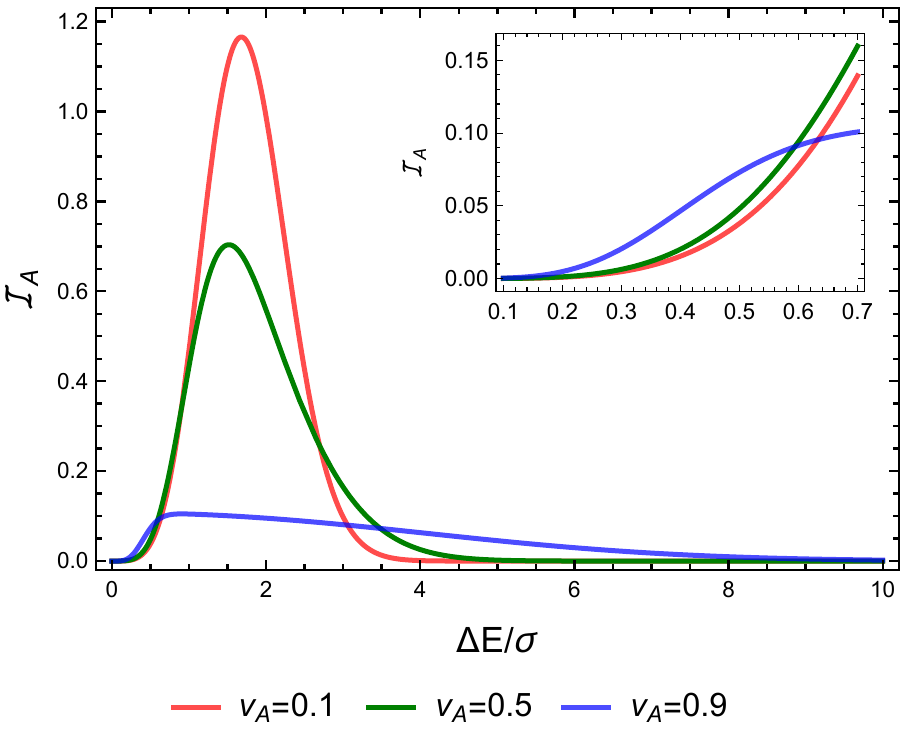}
\hskip 10pt
\includegraphics[width=0.42\textwidth]{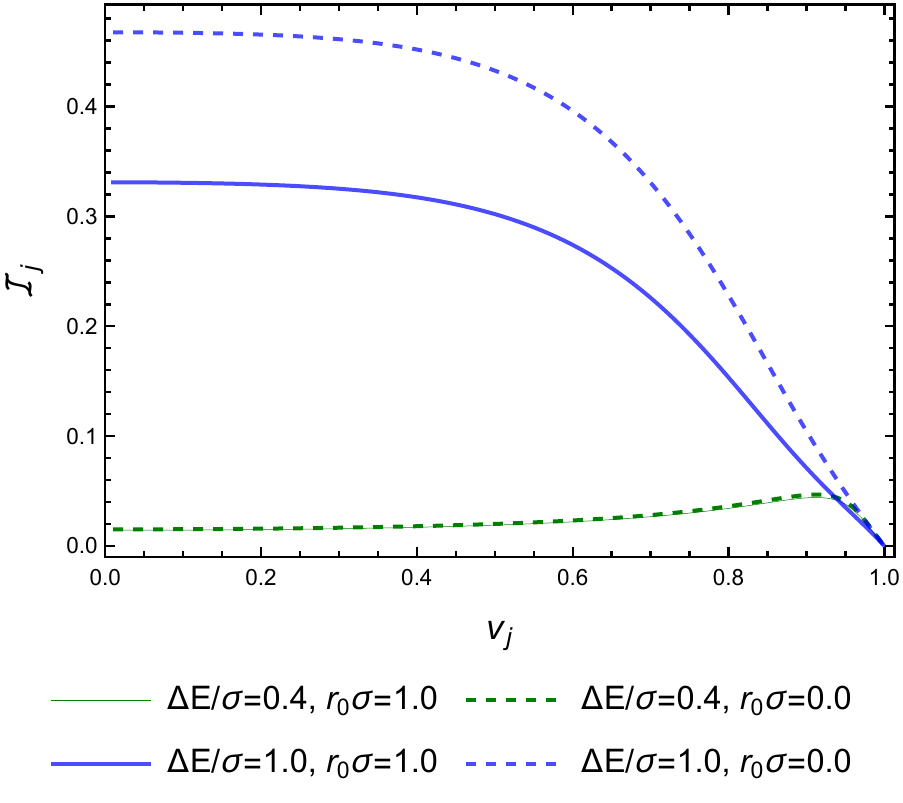}
\caption{In the upper figure $\mathcal{I}_{A}$ is plotted as a function of $\Delta 
E/\sigma$ for fixed values of $v_{A}$ in $(1+3)$ dimensions. On the other hand, in the lower figure $\mathcal{I}_{j}$ is 
plotted with respect to $v_{j}$ for different fixed values of $\Delta E/\sigma$ and $r_{0}\sigma$. 
In both of these figures we have considered the distribution function $f(\omega_{k})=C~\omega_{k} 
~e^{-(\omega_{k}-\omega_{0})^{2}/2\sigma^{2}}$ with $\omega_{0}/\sigma=0.5$. Here also
in the lower plot $r_{0}\sigma=1$ corresponds to detector $B$ and $r_{0}\sigma=0$ corresponds to 
detector $A$.}
\label{fig:Ij-1p3-GF}
\end{figure}

\subsubsection{Evaluation of $\mathcal{I}_{j}$}

Let us consider evaluating the integral $\mathcal{I}_{B}$ first, and one may expect that the 
expression of the integral $\mathcal{I}_{A}$ will arrive as a special case when $r_{0}=0$.
For trajectory given in (\ref{eq:ParaTrajecB}) the integral $\mathcal{I}_{B}$ can be evaluated 
considering the wave-vector in the spherical polar coordinates such that 
\begin{eqnarray}
k\cdot 
x&=&
-\omega_{k}t+\vec{k}.\vec{x}\no\\
&=&\omega_{k}(-\gamma_{B}\tau_{B}+x_{0}\sin\theta\cos\phi\no\\&&~~~~~~+y_{0}
\sin\theta\sin\phi+v_{B}\gamma_{B}\tau_{B}\cos\theta)~,
\end{eqnarray}
where we have used
\begin{eqnarray}
\vec{k} &=& \omega_{k}\left(\sin\theta\cos\phi\,\hat{x}+\sin\theta\sin\phi\,\hat{y}
+\cos\theta\,\hat{z}\right)~,\no\\
\vec{x} &=& x\,\hat{x}+y\,\hat{y}+z\,\hat{z}~,
\end{eqnarray}
with $\omega_{k}=|\vec{k}|$.
Then in this case one can obtain the quantity from Eq. (\ref{eq:Ijv-Ijnv-2a}) as
\begin{eqnarray}\label{eq:ADE-1p3-B}
\mathcal{A}_{B}(\Delta E)&=&
\int_{-1}^{1}\frac{du}{4\pi} \frac{\Delta E\,f\left(\frac{\Delta 
E}{\gamma_{B}(1-v_{B}u)}\right)}{\left[\gamma_{B}(1-v_{B}u)\right]^{2}}J_{0}\left(
\tfrac{r_{0}\Delta E\sqrt{1-u^{2}}}{\gamma(1-v_{B}u)}\right)~,\no\\
\end{eqnarray}
where $J_{n}(x)$ denotes the Bessel function of the first kind, and $r_{0}= \sqrt{x_{0}^{2} + 
y_{0}^{2}}$. A detailed derivation of this expression is presented in Appendix 
\ref{Appn:1p3-parallel-integrals}. From this expression one can easily find out the value of 
$\mathcal{A}(\Delta E)$ for detector $A$ by making $r_{0}=0$ as is also evident from the 
trajectories (\ref{eq:ParaTrajecA}) and (\ref{eq:ParaTrajecB}). When $r_{0}=0$ and $v_{B}\to v_{A}$ 
one may consider a change of variables $z=\Delta E/\gamma_{A}(1-v_{A}u)$ in the previous equation 
such that it reduces to 
\begin{equation}\label{eq:ADE-1p3-A}
 \mathcal{A}_{A}(\Delta E) = \frac{1}{4\pi v_{A}\gamma_{A}} \int_{\Delta E/D_{A}}^{D_{A}\Delta 
E}\,dz\,f(z)~.
\end{equation}
where we have used the fact that $J_{0}(0)=1$. This expression matches exactly with the one obtained 
explicitly for the detector $A$ with trajectory (\ref{eq:ParaTrajecA}), and is provided in Appendix 
\ref{Appn:1p3-parallel-integrals}. The relevant expression is derived in Eq. (\ref{IAgen}) of the 
Appendix. One should notice that in Eq. (\ref{eq:ADE-1p3-A}) and (\ref{eq:ADE-1p3-B}) we have 
included subscript $A$ and $B$ to signify the specific detectors. One may find out the respective 
$\mathcal{I}_{j}$ from $\mathcal{I}_{j} = |\mathcal{A}_{j}(\Delta E)|^2$ as here also 
$\mathcal{I}_{j}^{vac} =0 $ and $\mathcal{B}(\Delta E)=0$ (see Appendix 
\ref{Appn:1p3-parallel-integrals} for details).

In $(1+3)$ dimensions one may use the expressions of a few possible distribution functions as given 
in Appendix \ref{Appn:dist-fns-1p3} to obtain $\mathcal{I}_{j}$. 
In particular, in Appendix \ref{Appn:dist-fns-1p3} the explicit forms of the normalization constants 
$(C)$ in the exponential damping and the Gaussian distribution functions are obtained, which are 
essential for procuring a numerical value for the above mentioned integral.
Here also, these quantities $\mathcal{I}_{j}$ correspond to individual detector transition 
probabilities with a non-vacuum singly excited field state in the background. We have plotted this 
transition probability in Fig. \ref{fig:Ij-1p3-ED} and \ref{fig:Ij-1p3-GF} for the exponential 
decaying and Gaussian distribution functions, respectively. To be specific, we have plotted the 
quantity $\mathcal{I}_{B}$ with respect to the velocity $v_{B}$ (when $r_{0}\neq 0$), where the special case of $r_{0}=0$ 
produces the transition probability $\mathcal{I}_{A}$ with $v_{B}$ now becoming $v_{A}$. The curves 
in these plots have somewhat similar features to the ones from $(1+1)$ dimensions. Specifically,  
for low and high transition energy $\Delta E$, the transition probability increases with increasing 
velocity, and in the intermediate $\Delta E$ regimes this nature is reversed.
Also visible from these plots is that the transition probability decreases with increasing $r_{0}$.

\subsubsection{Evaluation of $\mathcal{I}_{\varepsilon}$, and $\mathcal{C}_\mathcal{I}$}

Let us now evaluate the integral $\mathcal{I}_{\varepsilon}$ for these two detectors which are in parallel inertial 
motion. In this scenario the quantities 
$\mathcal{I}_{\varepsilon}^{nv}$ and $\mathcal{I}_{\varepsilon_{W}}^{vac}$  
vanish. The detailed calculation is provided in Appendix \ref{Appn:1p3-parallel-integrals}. 
The only non-vanishing contribution is given by the integral $\mathcal{I}_{\varepsilon_{R}}^{vac}$. Using the representation from Eq. (\ref{eq:Ie-integral-b}) and with the help of Eqs. 
(\ref{eq:ParaTrajecA}) and (\ref{eq:ParaTrajecB}) one can express this integral as
\begin{eqnarray}\label{eq:1p3-parallel-IeVacR-1}
\mathcal{I}_{\varepsilon_{R}}^{vac} &=&
	-\intinf \intinf \frac{dp~dq}{2v_{A}v_{B}'\gamma_{A}\gamma_{B}} 
e^{i\Delta{E}(a_{1}p+a_{2}q)}\theta(a_{3}p-q)\no\\&&\times\int_{-1}^{1}du\intsinf 
\frac{\omega_{k}d\omega_{k}}{2(2 \pi)^{2} } 
\Big[e^{-i\omega_{k}a_{4}(a_{3}p-q)+i\omega_{k}\sqrt{p^{2}+ 
r_{0}^{2}}u}\no\\&&~~~~~~~~~~~~~~~~~~~~~~~~~-e^{i 
\omega_{k}a_{4}(a_{3}p-q)-i\omega_{k}\sqrt{p^{2}+r_{0}^{2}}u }\Big],\no\\
\end{eqnarray}
where we have considered the change of variables $p = v_{A}\gamma_{A}\tau_{A} - 
v_{B}\gamma_{B}\tau_{B}$ and $q = v_{A}\gamma_{A}\tau_{A} + v_{B}\gamma_{B}\tau_{B}$. The 
Jacobian of this transform is given by $|J| = 1/(2v_{A}v_{B}\gamma_{A}\gamma_{B})$. Whereas, the 
other variable is $u=\cos{\theta}$, and we have considered the redefined parameters $a_{1} = 
1/(2v_{A}\gamma_{A})-1/(2v_{B}\gamma_{B})$, $a_{2} = 1/(2v_{A}\gamma_{A}) + 1/(2v_{B}\gamma_{B})$, 
$a_{3} = (v_{A}+v_{B})/(v_{A}-v_{B})$, and $a_{4} = (v_{A}-v_{B})/2v_{A}v_{B}$. One may notice that 
due to the Heaviside step function the upper limit of the $q$ integration transforms to $a_{3}p$. 
Then one may evaluate this entire quantity $\mathcal{I}_{\varepsilon_{R}}^{vac}$ in a step by manner 
with first carrying out the integration over $q$. Subsequently the integrals over $u$ and 
$\omega_{k}$ can be performed in a likewise manner. After carrying out all these steps the  integral 
looks like
\begin{equation}\label{eq:1p3-parallel-IeVacR-2}
\mathcal{I}_{\varepsilon_{R}}^{vac} = \intinf  
dp\,\frac{i\,e^{i\Delta{E}p(a_{1}+a_{2}a_{3})}\cos\left(\tfrac{a_{2}\Delta{E}\sqrt{p^{2}+ 
r_{0}^{2}}}{a_{4}}\right)}{8\pi 
v_{A}v_{B}\gamma_{A}\gamma_{B} a_{4}\sqrt{p^{2}+r_{0}^{2}}}\,.
\end{equation}
To perceive a detailed procedure to arrive at this expression one may look into 
Appendix \ref{Appn:1p3-parallel-integrals-Ie-momentum}. This integral can be carried out to give 
a further simplified expression with the help of the integral representations of the modified 
Bessel function of the second kind $K_{n}(z)$ (see Eq. (3.876) of \cite{Gradshteyn}), which are
\begin{eqnarray}\label{eq:BesselK0-IntRep}
\int_{0}^{\infty} \frac{\cos \left(\varrho 
\sqrt{\xi^{2}+a^{2}}\right)}{\sqrt{\xi^{2}+a^{2}}} \cos (\beta \xi)
\mathrm{~d} \xi 
&=&K_{0}\left(a \sqrt{\beta^{2}-\varrho^{2}}\right)~;\no\\
\int_{0}^{\infty} \frac{\sin \left(\varrho 
\sqrt{\xi^{2}+a^{2}}\right)}{\sqrt{\xi^{2}+a^{2}}} \cos (\beta \xi) 
\mathrm{~d} \xi 
&=&0\,,~ {[\beta>\varrho>0] }\,.
\end{eqnarray}
Now one can observe that the exponential in the integral (\ref{eq:1p3-parallel-IeVacR-2}) can be 
written in terms of $\sin$ and $\cos$ functions so that the previous forms of Eq. 
(\ref{eq:BesselK0-IntRep}) can be perceived. In our case, a comparison with Eq. 
(\ref{eq:BesselK0-IntRep}) reveals $\beta = \Delta{E}(a_{1}+a_{2}a_{3})$ and $\varrho = 
a_{2}\Delta{E}/a_{4}$. Then from the consideration of a positive transition frequency $\Delta E$ 
(which is true in our case) and the explicit expressions of $a_{1}$, $a_{2}$, $a_{3}$, and $a_{4}$ 
one can confirm the satisfaction of the condition $\beta>\varrho>0$ even with the integral 
(\ref{eq:1p3-parallel-IeVacR-2}).
In particular, the integral of Eq. 
(\ref{eq:1p3-parallel-IeVacR-2}) now becomes
\begin{eqnarray}\label{eq:1p3parallel2}
\mathcal{I}_{\varepsilon_{R}}^{vac} &=& i\intsinf  
dp\,\frac{\cos\left(\frac{a_{2}\Delta{E}}{a_{4}}\sqrt{p^{2}+ 
r_{0}^{2}}\right)}{2\pi (v_{A}-v_{B})\gamma_{A}\gamma_{B}\sqrt{p^{2}+r_{0}^{2}}} 
\no\\
~&& ~~~~~~~~~~~~~~~\times~ 
\cos\left\{\Delta{E}p(a_{1}+a_{2}a_{3})\right\}\no\\&=&\frac{i~K_{0}\left(r_{0}\Delta{E} \sqrt { 
\left(a_{1}+a_{2}a_{3}\right)^ { 2
}-\left(a_{2}/a_{4}\right)^{2}}\right)}{2\pi (v_{A}-v_{B})\gamma_{A}\gamma_{B}}\no\\&=& 
\frac{-i}{2\pi\gamma_{A}\gamma_{B}(v_{A}-v_{B})}\no\\&&\times 
K_{0}\left(\tfrac{r_{0}\Delta{E}}{\gamma_{B}}\sqrt{\left(\tfrac{\frac{1}{\gamma_{A}}+\gamma_{B}(1-v_ 
{ A }v_{B})}{v_{A}-v_{B}}\right)^{2}-\gamma_{B}^{2}}\right),~~~~
\end{eqnarray}
where we have used (\ref{eq:BesselK0-IntRep}) along with the relation $2v_{A}v_{B}a_{4} = 
(v_{A}-v_{B})$.

\color{black}

As a side remark and for comparison with the earlier result (given in \cite{Koga:2018the}) we mention that in the limit 
$v_{B}\to0$ we have $(a_{1} + a_{2}a_{3})^{2} - (a_{2}/a_{4})^{2} = 
2(1+\gamma_{A})/v_{A}^{2}\gamma_{A}^{2}$, and thus the expression of $\mathcal{I}_{\varepsilon}$ 
(which is entirely determined by $\mathcal{I}_{\varepsilon_{R}}^{vac}$) is given by
\begin{eqnarray}\label{eq:KogaResult}
&&\left|\mathcal{I}_{\varepsilon}\right|=\frac{1}{2\pi 
v_{A}\gamma_{A}}K_{0}\left(\frac{r_{0}\Delta{E}}{v_{A}\gamma_{A}}\sqrt{2(1+\gamma_{A})}\right)~.
\end{eqnarray}
This result matches correctly with the one from \cite{Koga:2018the} (see Eq. (74) of 
\cite{Koga:2018the} with $\theta=\pi/2$, where $\theta$ is the angle between $\vec{v}$ and 
$\vec{x}_{0}$), where one of the detectors was taken to be static.

We should emphasize here that in our present case 
the quantities $\mathcal{I}_{\varepsilon_{W}}^{vac}$ and $\mathcal{I}_{\varepsilon}^{nv}$ vanish (see Appendix \ref{Appn:1p3-parallel-integrals} for detailed estimations of these 
integrals).
%
\begin{figure}[!h]
\centering
\includegraphics[width=0.42\textwidth]{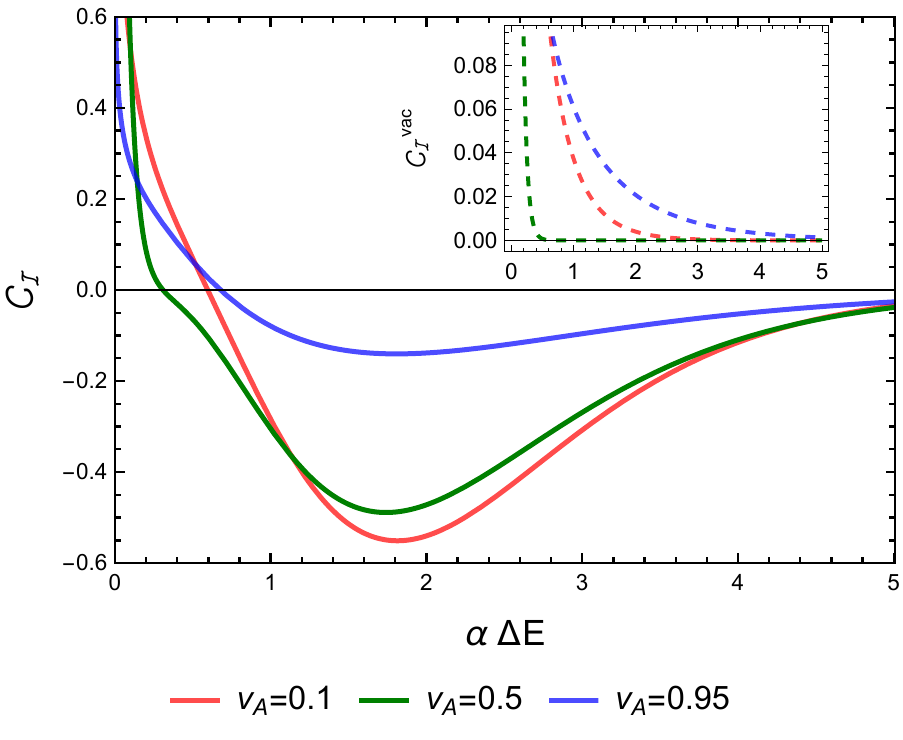}
\includegraphics[width=0.42\textwidth]{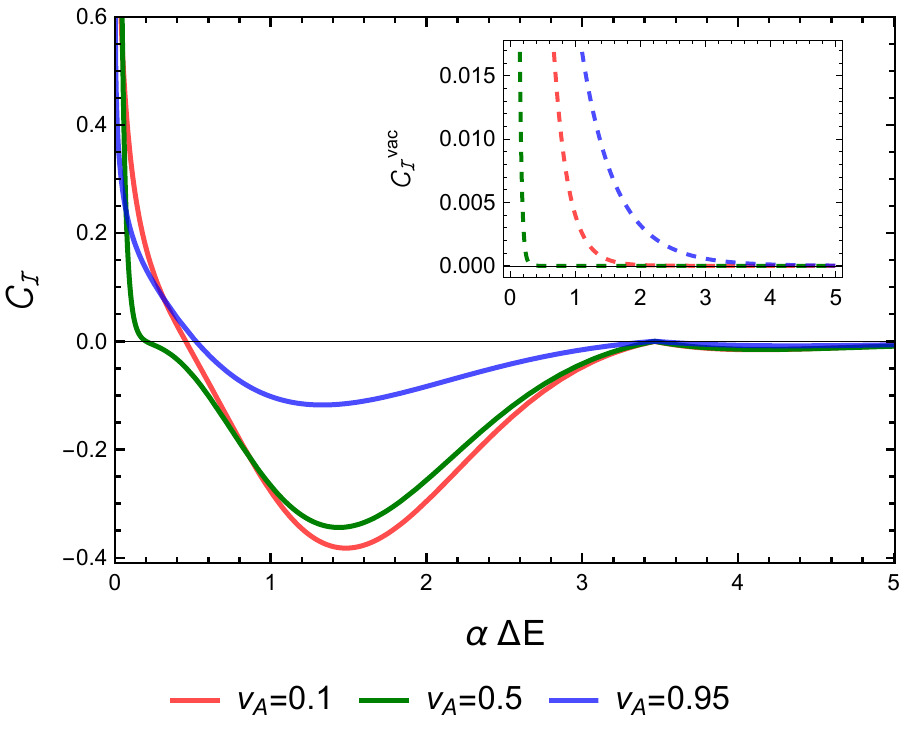}
\caption{In $(1+3)$ dimensions the integral $\mathcal{C}_{\mathcal{I}}$ signifying 
the concurrence is plotted as a function of $\alpha\,\Delta E$ for different fixed values of 
$v_{A}$ with two detectors in parallel inertial motion. In this plot we have considered the 
distribution function $f(\omega_{k})=C\, \omega_{k}\,e^{-\alpha\omega_{k}}$ for the singly excited 
background field state. Here the velocity of detector $B$ is fixed at $v_{B}=0.55$ and the other 
parameter is $r_{0}/\alpha = 0.5$ and $r_{0}/\alpha = 1$ respectively in the upper and lower plots. These curves assert that entanglement harvesting from single particle 
field state is possible in the low $\alpha\,\Delta E$ regimes. The inner plots correspond to 
$\mathcal{C}_{\mathcal{I}}^{vac}$, the concurrence, if the detectors were interacting with the field 
vacuum.}
\label{fig:C-1p3-para-vDE-ED}
\end{figure}
\begin{figure}[!h]
\centering
\includegraphics[width=0.42\textwidth]{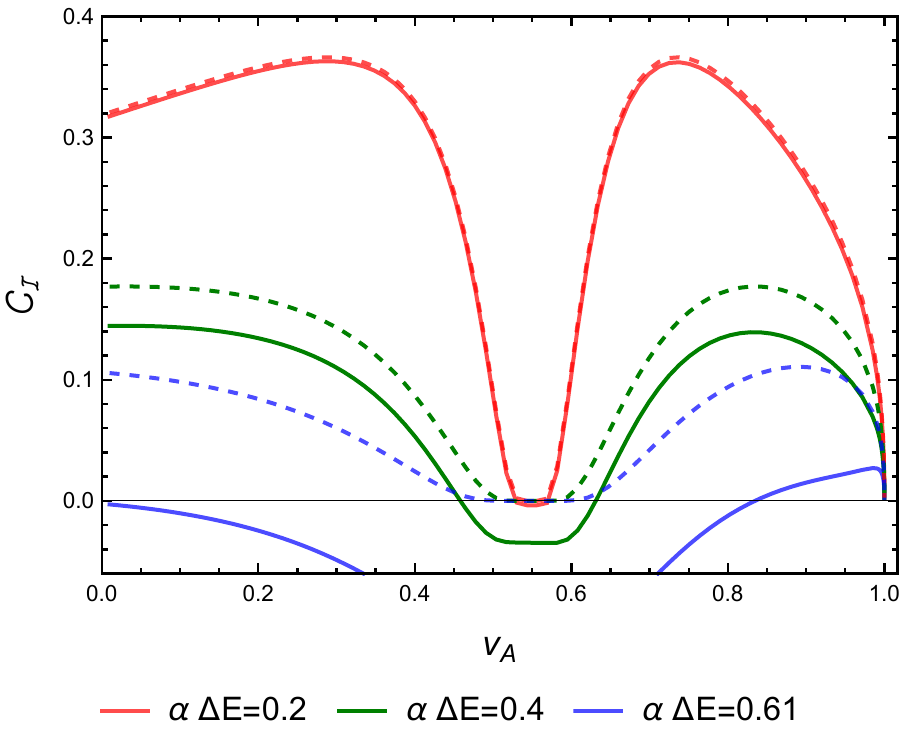}
\hskip 10pt
\includegraphics[width=0.42\textwidth]{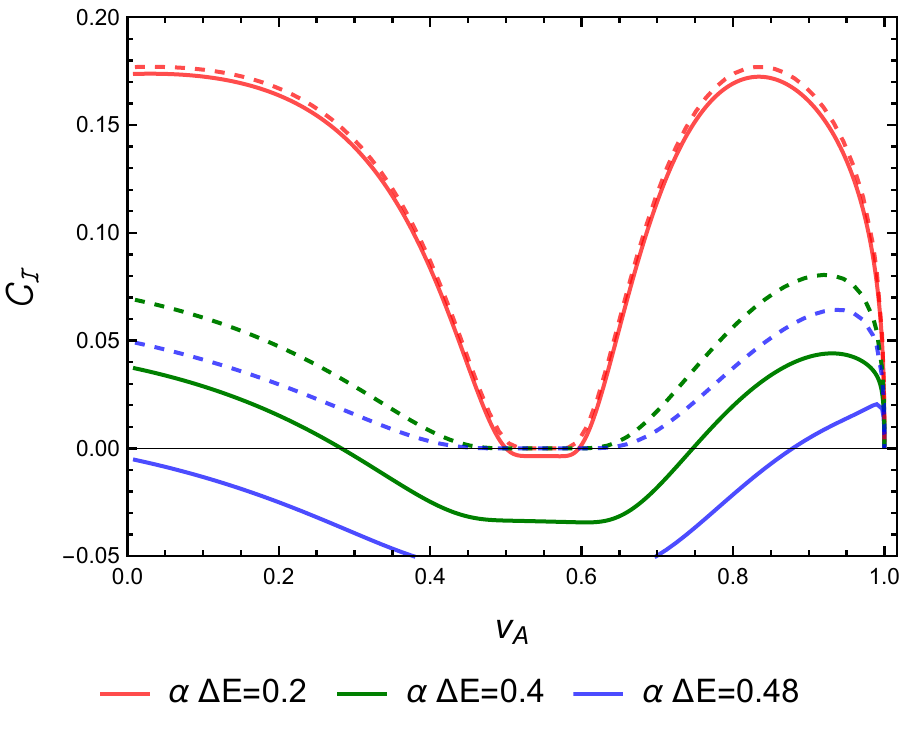}
\caption{In $(1+3)$ dimensions the integral $\mathcal{C}_{\mathcal{I}}$ signifying 
the concurrence is plotted as a function of $v_{A}$ for two detectors in parallel inertial motion. 
The upper and the lower plots respectively correspond to $r_{0}/\alpha = 0.5$ and $r_{0}/\alpha = 
1$. In both of these plots we have considered the distribution function $f(\omega_{k})=C\, 
\omega_{k}\,e^{-\alpha\omega_{k}}$ for the singly excited background field state. Here the velocity 
of detector $B$ is fixed at $v_{B}=0.55$. From these plots it is evident that with decreasing 
$\alpha\,\Delta E$ and $r_{0}/\alpha$ the concurrence increases. The solid lines denote the 
contributions from $\mathcal{C}_{\mathcal{I}}$, while the dashed lines denote 
$\mathcal{C}_{\mathcal{I}}^{vac}$.}
\label{fig:C-1p3-para-ED}
\end{figure}
\begin{figure}[!h]
\centering
\includegraphics[width=0.42\textwidth]{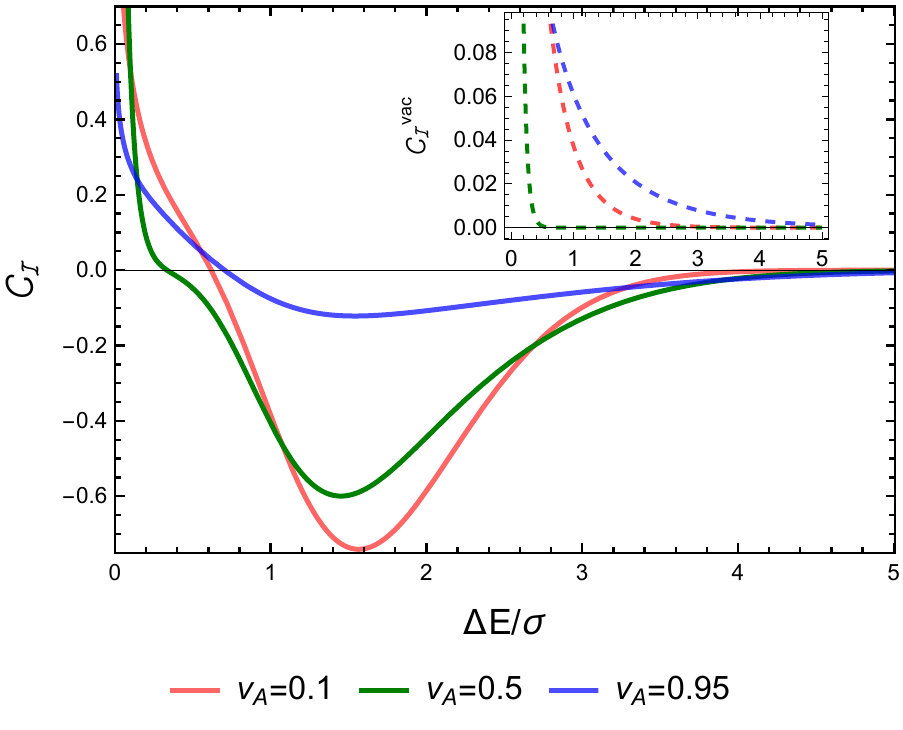}
\includegraphics[width=0.42\textwidth]{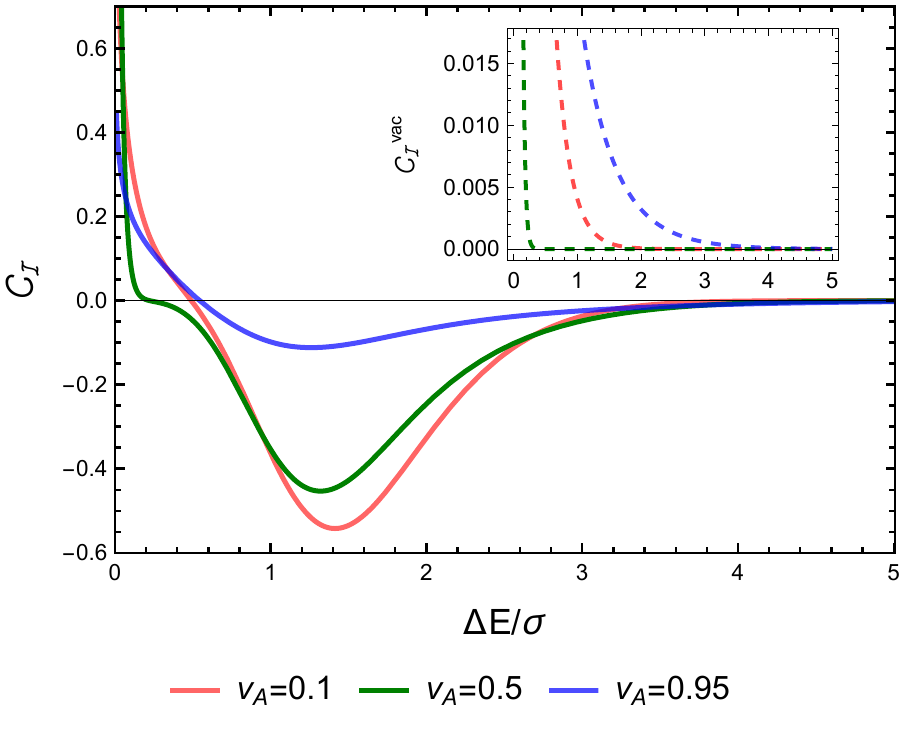}
\caption{In $(1+3)$ dimensions the integral $\mathcal{C}_{\mathcal{I}}$ signifying 
the concurrence is plotted as a function of $\Delta E/\sigma$ for different fixed values of $v_{A}$ 
with two detectors in parallel inertial motion. In this plot we have considered the distribution 
function $f(\omega_{k})=C\, \omega_{k} ~e^{-(\omega_{k}-\omega_{0})^{2}/2\sigma^{2}}$ for the singly 
excited background field state. Here the velocity of detector $B$ is fixed at $v_{B}=0.55$ and the 
other fixed parameters are $r_{0}\sigma = 0.5$ and $r_{0}\sigma = 1$ (respectively for the upper and lower figures), $\omega_{0}/\sigma=0.5$. Here also one can observed 
that entanglement harvesting from single particle field state is possible in the low $\Delta 
E/\sigma$ regimes. The inner plots correspond to $\mathcal{C}_{\mathcal{I}}^{vac}$, the concurrence, 
if the detectors were interacting with the field vacuum.}
\label{fig:C-1p3-para-vDE-GF}
\end{figure}
\begin{figure}[!h]
\includegraphics[width=0.42\textwidth]{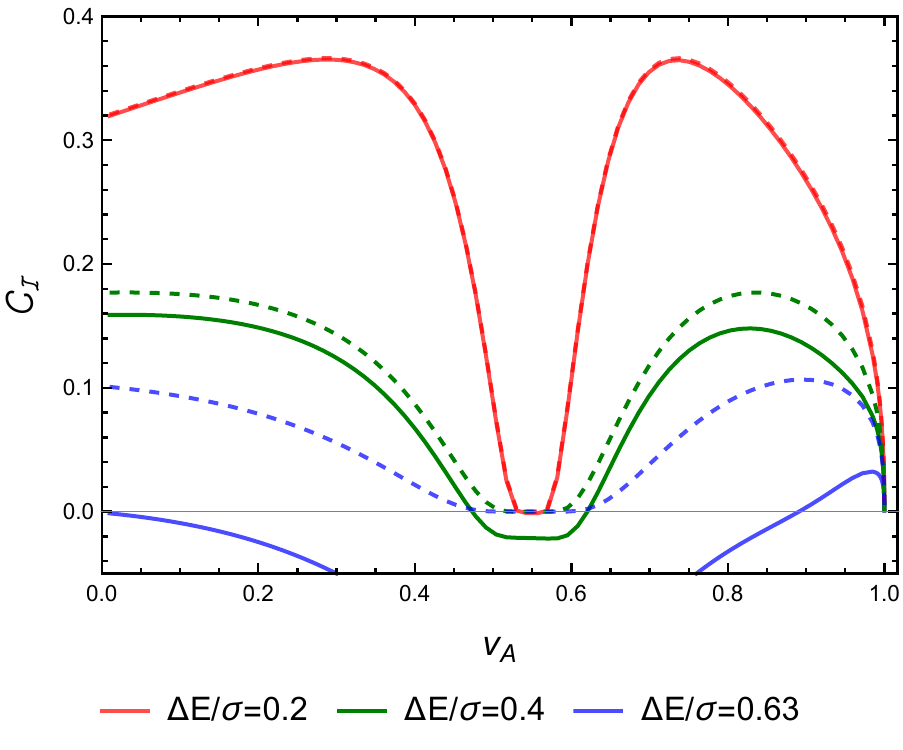}
\hskip 10pt
\includegraphics[width=0.42\textwidth]{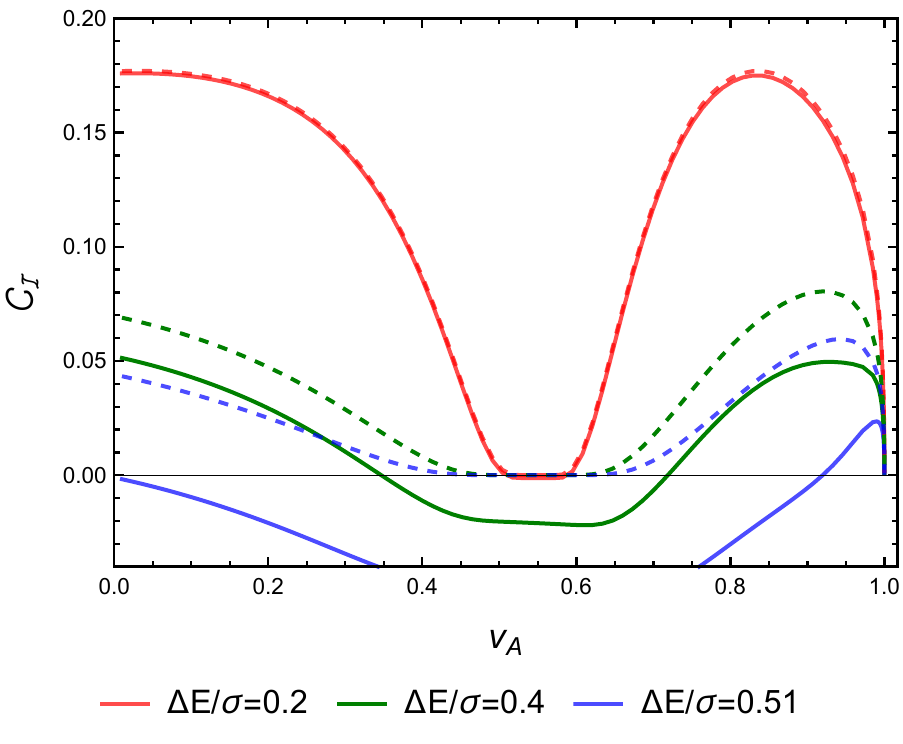}
\caption{In $(1+3)$ dimensions the integral $\mathcal{C}_{\mathcal{I}}$ signifying 
the concurrence is plotted as a function of $v_{A}$ for two detectors in parallel inertial motion. 
The upper and the lower plots respectively correspond to $r_{0}\,\sigma = 0.5$ and $r_{0}\,\sigma = 
1$. In both of these plots we have considered the distribution function $f(\omega_{k})=C\, 
\omega_{k} ~e^{-(\omega_{k}-\omega_{0})^{2}/2\sigma^{2}}$ for the singly excited background field 
state. Here the velocity of detector $B$ is fixed at $v_{B}=0.55$, and the other parameter 
$\omega_{0}/\sigma=0.5$. From these 
plots it is evident that with decreasing $\Delta E/\sigma$ and $r_{0}\,\sigma$ the concurrence 
increases. The solid lines denote the contributions from $\mathcal{C}_{\mathcal{I}}$, while the 
dashed lines denote $\mathcal{C}_{\mathcal{I}}^{vac}$.} 
\label{fig:C-1p3-para-GF}
\end{figure}
%
Here one may notice that in the limit of $(v_{A}-v_{B})\to 0$ Eq. (\ref{eq:1p3parallel2}) leads to 
vanishing result. Therefore, for two parallel co-moving detectors the quantity dictating 
entanglement harvesting vanishes and one can assert that in this situation there will be no 
entanglement harvesting. A similar conclusion can also be made starting from Eq. 
(\ref{eq:KogaResult}), where one of the two observers is static, that the quantity 
$\left|\mathcal{I}_{\varepsilon}\right|$ vanishes in the limit $v_{A}\to 0$.

In Fig. \ref{fig:C-1p3-para-vDE-ED}, \ref{fig:C-1p3-para-ED}, \ref{fig:C-1p3-para-vDE-GF}, and 
\ref{fig:C-1p3-para-GF} we have plotted the concurrence denoting quantity 
$\mathcal{C}_{\mathcal{I}}$ for different distribution functions of the single particle field state.
One can notice from these figures (specifically Fig. 
\ref{fig:C-1p3-para-vDE-ED} and \ref{fig:C-1p3-para-vDE-GF}) that with decreasing $\Delta E$ the concurrence quantitatively 
increases, signifying an increase in the measure of the harvested entanglement.
While from Fig. 
\ref{fig:C-1p3-para-ED} and \ref{fig:C-1p3-para-GF} it is evident that when $v_{A}\to v_{B}$ the entanglement harvesting vanishes. 
Moreover, when the detector transition energy $\Delta E$ becomes larger we observe a wide region of 
no harvesting around $v_{A} = v_{B}$ and this increases with the increase of $\Delta E$.
When the $\Delta E$ value is very large the entanglement harvesting is happening only in high velocity regimes.
From Eq. (\ref{eq:KogaResult}) one can observe that in the limit $v_{A}\to 1$ the 
quantity $\left|\mathcal{I}_{\varepsilon}\right|$ vanishes, and this feature is apparent from these 
figures too.
One should also note that if the detectors were interacting with the field vacuum rather than the 
non-vacuum state, the quantity $\mathcal{I}_{j}$ would have been zero and the concurrence would have 
been completely given by $\mathcal{C}_{\mathcal{I}}^{vac} = \left|\mathcal{I}_{\varepsilon}\right|$. 
We have included the plots of these $\mathcal{C}_{\mathcal{I}}^{vac}$ in the above mentioned 
figures. 
%
%
Here also, like the $(1+1)$ dimensional case, the entanglement harvesting is suppressed due to 
non-vacuum fluctuations as compared to vacuum effect. Here both the magnitude of concurrence and 
valid region of the value of $v_A$ for entanglement harvesting decreases. From these figures we also observe that entanglement harvesting is 
decreasing with increasing distance $r_{0}/\alpha$ between the parallel paths, and with increasing 
detector transition energy $\Delta E$.

\color{black}

\subsection{Perpendicular motion}\label{subsec:1p3D-perpendicular-UV}

In this part we consider two Unruh-DeWitt detectors in motion perpendicular to each other. We 
consider detector $A$ with an uniform velocity along the $x$-axis and detector $B$ with an uniform velocity along the $z$ direction 
starting from a constant displacement in the $y$ direction. The trajectory of detector $A$ is
\begin{equation}\label{eq:PerpTrajecA}
t_{A}=\gamma_{A} \tau_{A};~~x_{A}=\gamma_{A} v_{A}\tau_{A};~~;~~y_{A}=0;~~z_{A}=0~.
\end{equation}
On the other hand, the trajectory of detector $B$ is given by
\begin{equation}\label{eq:PerpTrajecB}
t_{B}=\gamma_{B} \tau_{B};~~x_{B}=0;~~y_{B}=r_{0};~~z=\gamma_{B} v_{B}\tau_{B}~.
\end{equation}
These trajectories signify that the motion of the detectors are confined to the $x-z$ plane, while 
there is a perpendicular initial separation between them in the $y$ direction.


\subsubsection{Evaluation of $\mathcal{I}_{j}$}
One may observe that the trajectory of the detector $A$ in this case and the previous case of the 
parallel motion are physically the same. Therefore, the expressions of the integral 
$\mathcal{I}_{A}$ will also be the same. In fact this indeed comes to be true and one can look into 
Appendix \ref{Appn:1p3-perpendicular-integrals-IA} for a derivation of the expression of 
$\mathcal{I}_{A}$ with the trajectory (\ref{eq:PerpTrajecA}). Here we recall the expression of 
$\mathcal{A}_{A}(\Delta E)$ from Eq. (\ref{eq:ADE-1p3-A}), which is relevant to estimate 
$\mathcal{I}_{A}$, as
\begin{eqnarray}\label{eq:ADE-1p3-A-pp}
 \mathcal{A}_{A}(\Delta E) = \frac{1}{4\pi v_{A}\gamma_{A}} \int_{\Delta E/D_{A}}^{D_{A}\Delta 
E}\,dz\,f(z)~.
\end{eqnarray}
On the other hand, in a similar manner one can also find out the quantity $\mathcal{A}_{B}(\Delta 
E)$ relevant for estimating $\mathcal{I}_{B}$ with the trajectory (\ref{eq:PerpTrajecB}) as
\begin{eqnarray}
\mathcal{A}_{B}(\Delta 
E) &=& \int_{-1}^{1}\frac{du}{4\pi} \frac{\Delta E~f\left(\frac{\Delta 
E}{\gamma_{B}(1-v_{B}u)}\right)}{[\gamma_{B}(1-v_{B}u)]^{2}}J_{0}\left( \tfrac{r_{0}\Delta 
E\sqrt{1-u^{2}}}{\gamma_{B}(1-v_{B}u)}\right)\,.\no\\
\end{eqnarray}
This expression can be obtained from Eq. (\ref{eq:ADE-1p3-B}) as the trajectories of 
detector $B$ from the parallel and perpendicular motion cases are physically equivalent. In this 
regard, compare the trajectories of detector $B$ from Eq. (\ref{eq:ParaTrajecB}) and 
(\ref{eq:PerpTrajecB}). Furthermore, one can find that here also $\mathcal{I}_{j}^{vac} = 0 $ and 
$\mathcal{B}_{j}(\Delta E) = 0$, see Appendix \ref{Appn:1p3-perpendicular-integrals} for details. 
Then the previous integral signifying individual detector transition probability for the detectors 
$A$ and $B$ are given by $\mathcal{I}_{j} = |\mathcal{A}_{j}(\Delta E)|^2$.

Here with the two detectors in inertial motion perpendicular to each other in $(1+3)$ dimensions, 
the integrals $\mathcal{I}_{j}$ are the same as the previous parallel case (see Eq. 
(\ref{eq:ADE-1p3-A}) and (\ref{eq:ADE-1p3-B}) and compare them with the expressions here). Therefore 
the individual detector's response function will have the same features as shown in Fig. 
\ref{fig:Ij-1p3-ED} and \ref{fig:Ij-1p3-GF}.

\subsubsection{Evaluation of $\mathcal{I}_{\varepsilon}$, and $\mathcal{C}_\mathcal{I}$}

Now we shall evaluate the quantity $\mathcal{I}_{\varepsilon}$ for two detectors in perpendicular 
inertial motion with trajectories given in (\ref{eq:PerpTrajecA}) and (\ref{eq:PerpTrajecB}). In 
particular, here also one can observe that the $\mathcal{I}_{\varepsilon}^{nv}$ part of this 
quantity vanishes, so does the $\mathcal{I}_{\varepsilon_{W}}^{vac}$. Then 
$\mathcal{I}_{\varepsilon} = \mathcal{I}_{\varepsilon}^{nv} +\mathcal{I}_{\varepsilon_{W}}^{vac} + 
\mathcal{I}_{\varepsilon_{R}}^{vac}$ is completely given by $\mathcal{I}_{\varepsilon_{R}}^{vac}$. 
In Eq. (\ref{eq:Ie-integral-b}) we have seen that this integral is dependent on the retarded 
Green's function $G_{R}\left(X_{A},X_{B}\right)$. 
%
%
This retarded Green function $G_{R}\left(X_{A},X_{B}\right)$ for a massless, minimally coupled, 
free scalar field in the Minkowski spacetime (see \cite{Koga:2018the}) is expressed as
\begin{eqnarray}
G_{R}\left(X_{A},X_{B}\right) &=& -
\frac{1}{2 \pi}~\Theta
\left(t_{A}-t_{B}\right)\no\\
~&&~~\times~ \delta
\left(\left(t_{A}-t_{B}\right)^{2}-\left|\boldsymbol{x}_{A}-\boldsymbol{x}_{B}\right|^{2}\right)
\no\\&=&-\frac{1}{2 \pi} \Theta\left(t_{A}-t_{B}\right) 
\delta(g(t_{A},t_{B}))~.
\end{eqnarray}
For the considered trajectories of detector $A$ and $B$ from Eq. (\ref{eq:PerpTrajecA}) and 
(\ref{eq:PerpTrajecB}) one can find the solution of $ 
g(t_{A},t_{B})=t_{A}^{2}(1-v_{A}^{2})+t_{B}^{2}(1-v_{B}^{2})-2 t_{A} t_{B}-r_{0}^{2}=0$
with respect to $t_{B}$ as
\begin{equation}\label{MinGDeno}
t_{B}=\gamma_{B}^{2}(t_{A}\pm u(t_{A}))\equiv t_{\pm}~,
\end{equation}
with,
\begin{equation}
u(t_{A})=\sqrt{t_{A}^{2}(v_{A}^{2}+v_{B}^{2}-v_{A}^{2}v_{B}^{2})+r_{0}^{2}(1-v_{B}^{2})}~,
\end{equation}
where, for $0\le v_{j}\le1$ one can confirm that,
\begin{eqnarray}
&&\gamma_{B}^{2}(t_{A}+ u(t_{A}))>t_{A}\no\\&&\gamma_{B}^{2}(t_{A}- u(t_{A}))<t_{A}~.
\end{eqnarray}
Now it is imperative to use the property of Dirac delta distributions
\begin{eqnarray}
\delta(g(t_{A},t_{B}))&=&\left[\frac{\delta(t_{B}-t_{+})}{|g^{\p}(t_{A},t_{B})|_{t_{B}=t_{+}}}+\frac
{\delta(t_{B}-t_{-})}{|g^{\p}(t_{A},t_{B})|_{t_{B}=t_{-}}}\right]\no\\&=&\frac{1}{2u(t_{A})}[
\delta(t_{B}-t_{+})+\delta(t_{B}-t_{-})]~.
\end{eqnarray}
Then the integral $\mathcal{I}_{\varepsilon_{R}}^{vac}$ from Eq. 
(\ref{eq:Ie-integral-b}), with a change of variable from $\tau_{j}$ to $t_{j}$, can be 
expressed as
\begin{eqnarray}\label{eq:1p3-IeVacR}	
\mathcal{I}_{\varepsilon_{R}}^{vac} &=& \frac{-i}{2\pi}\intinf{\frac{dt_{A}}{\gamma_{A}}}\int_{
-\infty 
}^{t_{A}}{\frac{dt_{B}}{\gamma_{B}}}\scalebox{.9}{$e^{i\Delta{E}\big(\frac{t_{A}}{\gamma_{A}}+\frac{ 
t_{B}}{\gamma_{B}}\big)}\delta(g(t_{A},t_{B}))$}\no\\
&=& \frac{-i}{4\pi}\intinf{\frac{dt_{A}}{\gamma_{A}\gamma_{B} 
~u(t_{A})}}e^{i\Delta{E}\left(\frac{t_{A}}{\gamma_{A}}+\gamma_{B}(t_{A}-u(t_{A})\right)}
\no\\
&=&\frac{-i}{2\pi 
\gamma_{A}\gamma_{B}}\intsinf\scalebox{.95}{$\frac{dt_{A}e^{-i\Delta{E}\gamma_{B}u(t_{A})}}{u(t_{A})
}\cos$}\scalebox{.85}{$\left\{\Delta{E}t_{A}\left(\frac{1}{\gamma_{A}}+\gamma_{B}\right)\right\}$}
~.\no\\
\end{eqnarray}
Furthermore, one can expand the exponential in this integral in terms of $\sin$ and $\cos$ 
functions and then by comparing them with the integral representations of the Bessel functions 
(\ref{eq:BesselK0-IntRep}), can recognize 
\begin{eqnarray}
a=r_{0}/\gamma_{B};~~ p=\gamma_{B}\Delta{E};~~b=\frac{\Delta{E}\left(\frac{1}{\gamma_{A}}+\gamma_{B}\right)}{\sqrt{v_{A}^{2}+v_{B}^{2}(1-v_{A}^{2})}};
\end{eqnarray}
and as one has $0\le v_{j}\le1$, this also satisfies the condition among the parameters from 
(\ref{eq:BesselK0-IntRep}), which is $ b>p~$. Thus, the integration in Eq. (\ref{eq:1p3-IeVacR}) 
can be evaluated. One can provide the simplified form as
\begin{eqnarray}
\mathcal{I}_{\varepsilon_{R}}^{vac} &=& 
\frac{-i}{2\pi\gamma_{A}\gamma_{B}\sqrt{v_{A}^{2}+v_{B}^{2} (1-v_{A}^{2})}}\no\\&&\times 
K_{0}\left(r_{0}\Delta{E}\sqrt{\frac{\left(\frac{1}{\gamma_{A}\gamma_{B}}+1\right)^{2}}{v_{A}^{2}+v_
{B}^{2}(1-v_{A}^{2})}-1}\right)\,.
\end{eqnarray}
Moreover, one can check that when $v_{B}=0$ (in that situation $\gamma_{B}=1$), this expression 
reduces to (\ref{eq:KogaResult}) as obtained in the parallel case. Here also the quantities 
$\mathcal{I}_{\varepsilon_{W}}^{vac}$ and $\mathcal{I}_{\varepsilon}^{nv}$ vanish and we refer the 
reader to go through Appendix \ref{Appn:1p3-perpendicular-integrals} for detailed estimations of 
these integrals.
\begin{figure}[!h]
\centering
\includegraphics[width=0.42\textwidth]{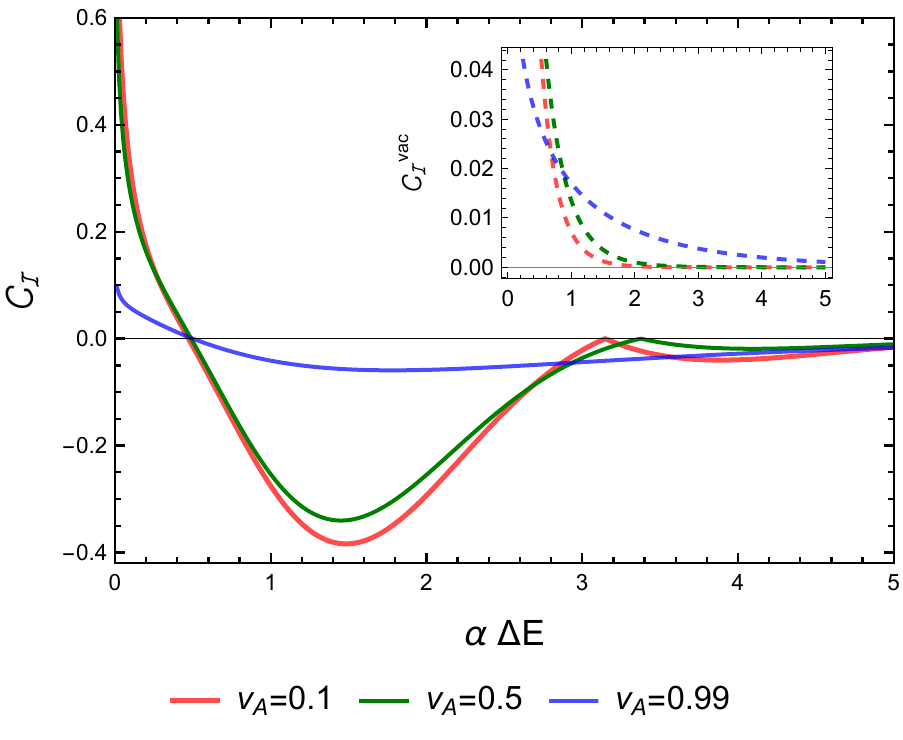}
\caption{In $(1+3)$ dimensions for perpendicular motion of the detectors the integral $\mathcal{C}_{\mathcal{I}}$ signifying 
the concurrence is plotted as a function of $\alpha\,\Delta E$ for different fixed values of 
$v_{A}$. In this plot we have considered the 
distribution function $f(\omega_{k})=C\, \omega_{k}\,e^{-\alpha\omega_{k}}$ for the singly excited 
background field state. Here the velocity of detector $B$ is fixed at $v_{B}=0.55$ and the other 
parameter is $r_{0}/\alpha = 1$. These curves assert that entanglement harvesting from single particle 
field state is possible in the low $\alpha\,\Delta E$ regimes. The inner plot correspond to 
$\mathcal{C}_{\mathcal{I}}^{vac}$, the concurrence, if the detectors were interacting with the field 
vacuum.  Similar nature is also obtained for $r_{0}/\alpha=0.5$ case.}
\label{fig:C-1p3-perp-vDE-ED}
\end{figure}
\begin{figure}[!h]
\centering
\includegraphics[width=0.42\textwidth]{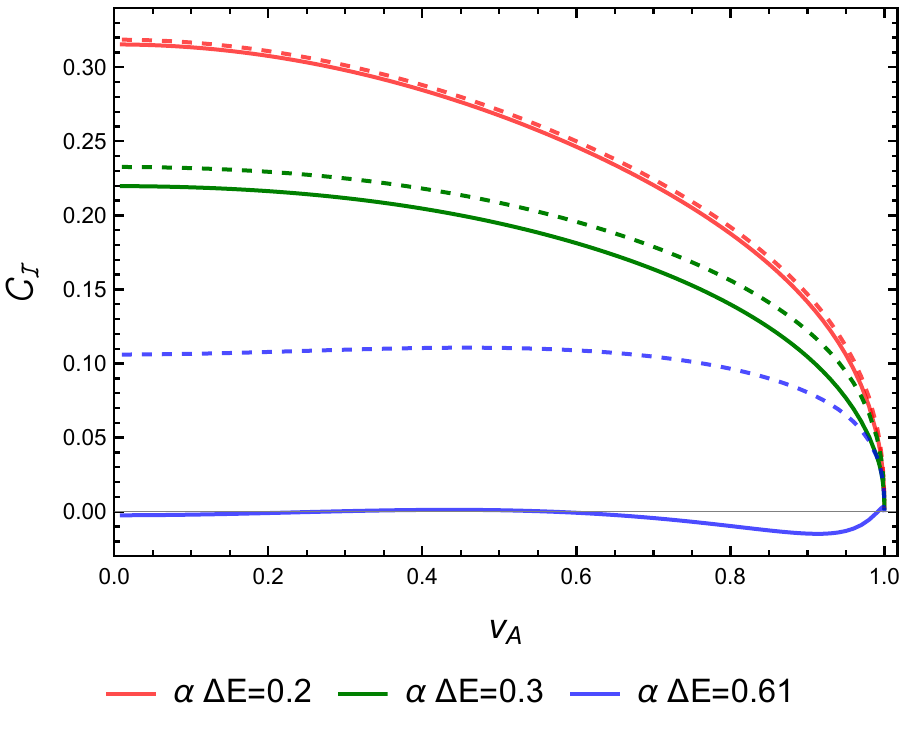}
\hskip 10pt
\includegraphics[width=0.42\textwidth]{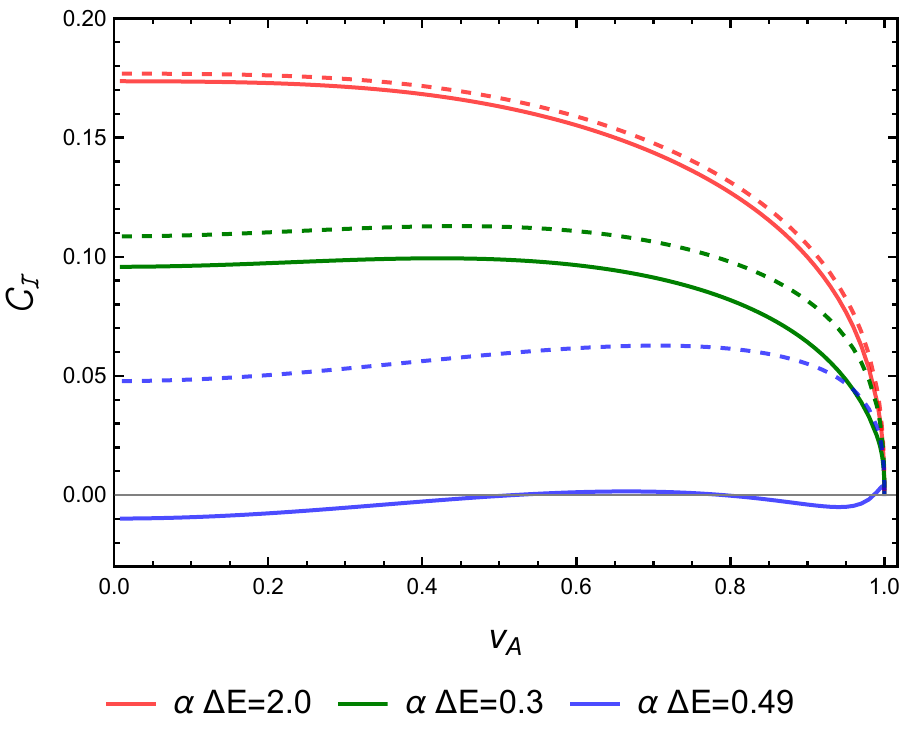}
\caption{In $(1+3)$ dimensions  for the detectors' perpendicular motion the 
integral $\mathcal{C}_{\mathcal{I}}$, signifying the concurrence, is plotted as a function of 
$v_{A}$. The upper and the lower plots respectively correspond to $r_{0}/\alpha = 0.5$ and 
$r_{0}/\alpha = 1$. In both of these plots we have considered the distribution function 
$f(\omega_{k})=C\, \omega_{k}\,e^{-\alpha\omega_{k}}$ for the singly excited background field state. 
Here the velocity of detector $B$ is fixed at $v_{B}=0.55$. From these plots it is evident that with 
decreasing $\alpha\,\Delta E$ and $r_{0}/\alpha$ the concurrence increases. The solid lines denote 
the contributions from $\mathcal{C}_{\mathcal{I}}$, while the dashed lines denote 
$\mathcal{C}_{\mathcal{I}}^{vac}$.}
\label{fig:C-1p3-perp-ED}
\end{figure}
\begin{figure}[!h]
\centering
\includegraphics[width=0.42\textwidth]{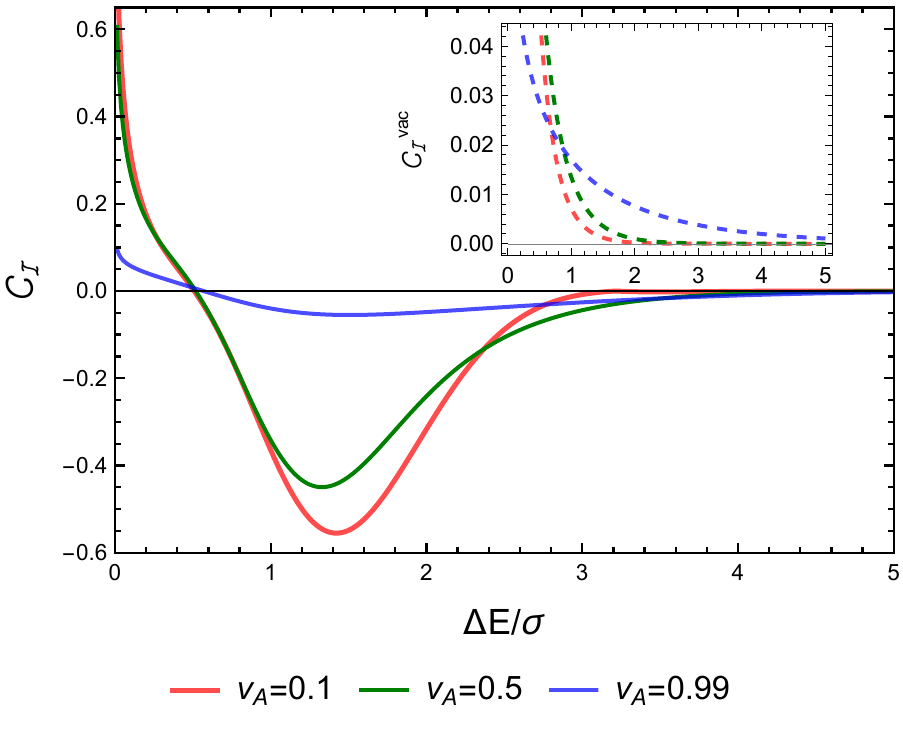}
\caption{In $(1+3)$ dimensions for perpendicular motion of the
detectors the integral $\mathcal{C}_{\mathcal{I}}$ signifying 
the concurrence is plotted as a function of $\Delta E/\sigma$ for different fixed values of $v_{A}$. In this plot we have considered the distribution 
function $f(\omega_{k})=C\, \omega_{k} ~e^{-(\omega_{k}-\omega_{0})^{2}/2\sigma^{2}}$ for the singly 
excited background field state. Here the velocity of detector $B$ is fixed at $v_{B}=0.55$ and the 
other fixed parameters are  $r_{0}\sigma = 1$ and $\omega_{0}/\sigma=0.5$. Here also one can observed 
that entanglement harvesting from single particle field state is possible in the low $\Delta 
E/\sigma$ regimes. The inner plot correspond to $\mathcal{C}_{\mathcal{I}}^{vac}$, the concurrence, 
if the detectors were interacting with the field vacuum. Similar nature is also obtained for $r_{0}\sigma=0.5$ case.}
\label{fig:C-1p3-perp-vDE-GF}
\end{figure}

\begin{figure}[!h]
\centering
\includegraphics[width=0.42\textwidth]{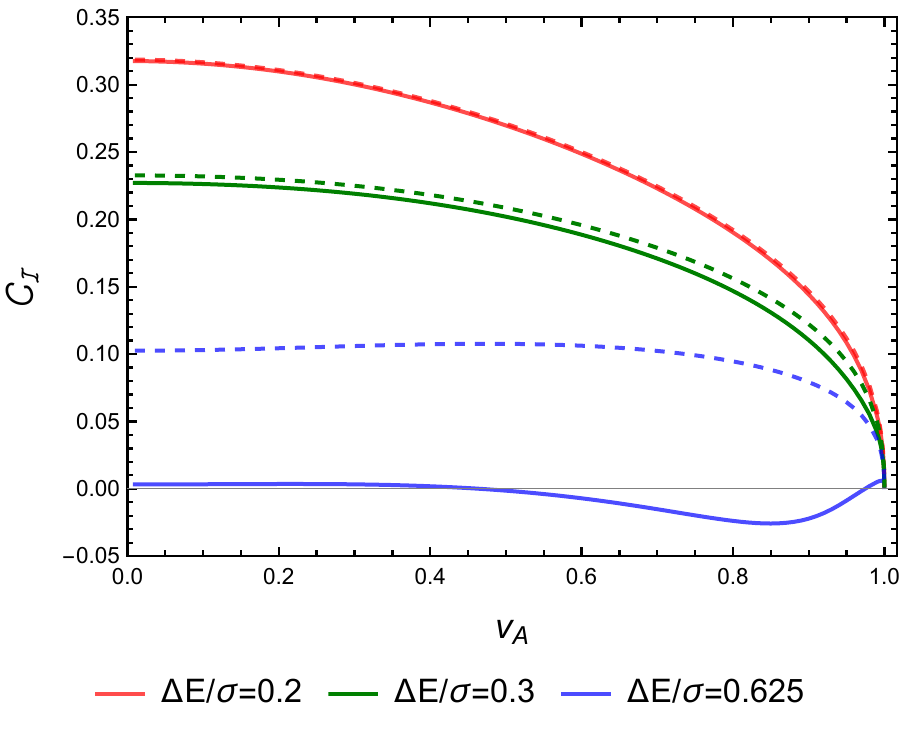}
\hskip 10pt
\includegraphics[width=0.42\textwidth]{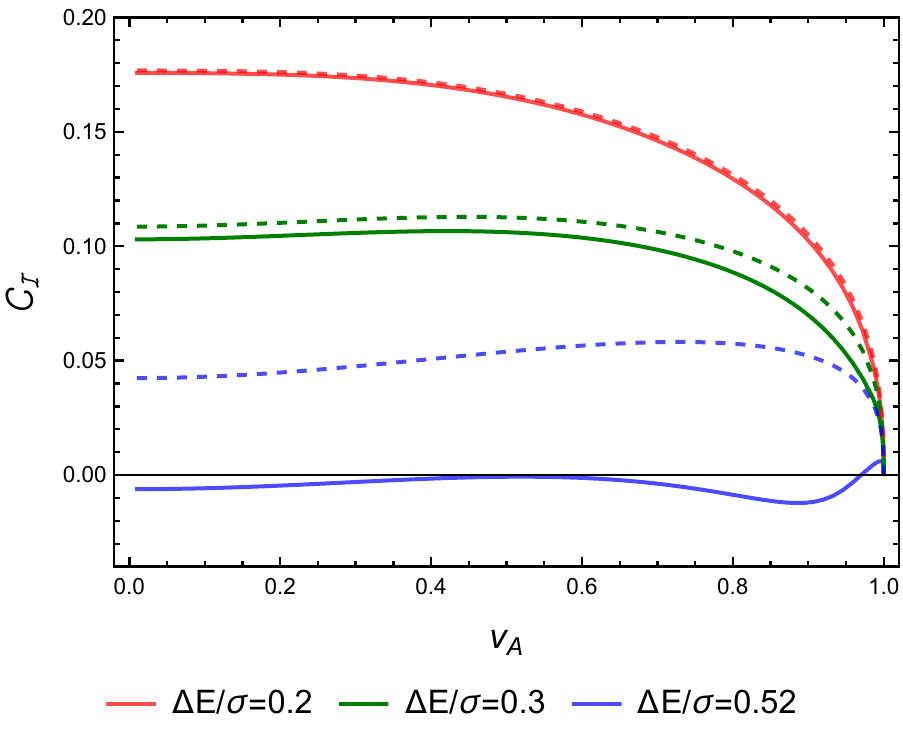}
\caption{In $(1+3)$ dimensions for the detectors in perpendicular inertial motion, 
the integral $\mathcal{C}_{\mathcal{I}}$ signifying the concurrence is plotted as a function of 
$v_{A}$. The upper and the lower plots respectively correspond to $r_{0}/\sigma = 0.5$ and 
$r_{0}/\sigma = 1$. In both of these plots we have considered the distribution function 
$f(\omega_{k})=C~\omega_{k} ~e^{-(\omega_{k}-\omega_{0})^{2}/2\sigma^{2}}$. Here the velocity of detector $B$ is fixed at $v_{B}=0.55$, and the other 
fixed parameter is $\omega_{0}/\sigma=0.5$. From these plots it is evident that with decreasing 
$\Delta E/\sigma$ and $r_{0}\,\sigma$ the concurrence increases. The solid lines denote the 
contributions from $\mathcal{C}_{\mathcal{I}}$, while the dashed lines denote 
$\mathcal{C}_{\mathcal{I}}^{vac}$.}
\label{fig:C-1p3-perp-GF}
\end{figure}
\begin{figure}[!h]
\centering
\includegraphics[width=0.42\textwidth]{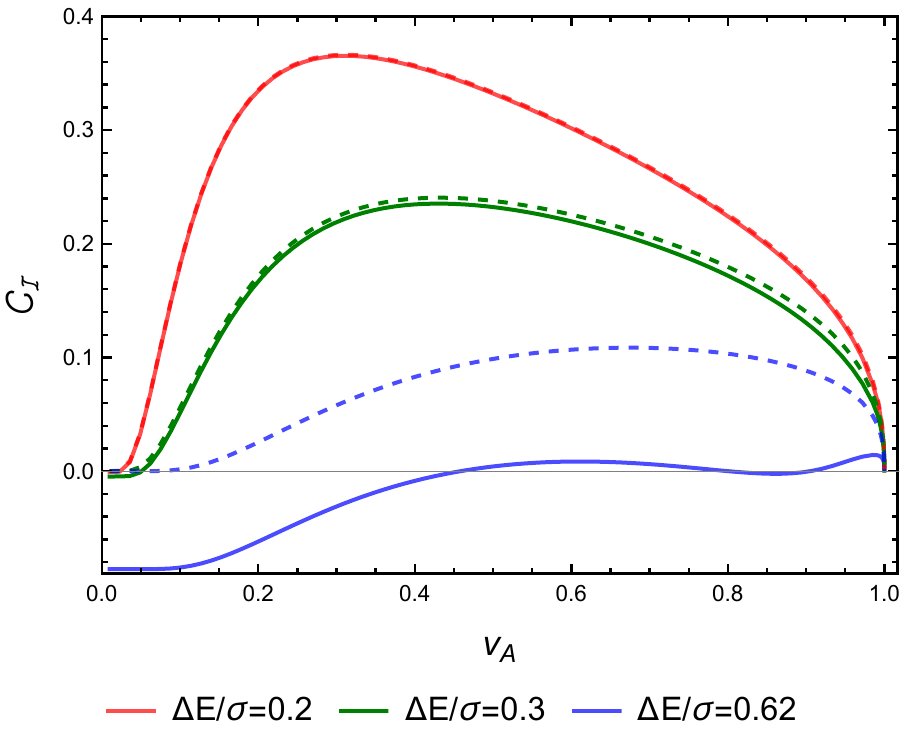}
\hskip 10pt
\includegraphics[width=0.42\textwidth]{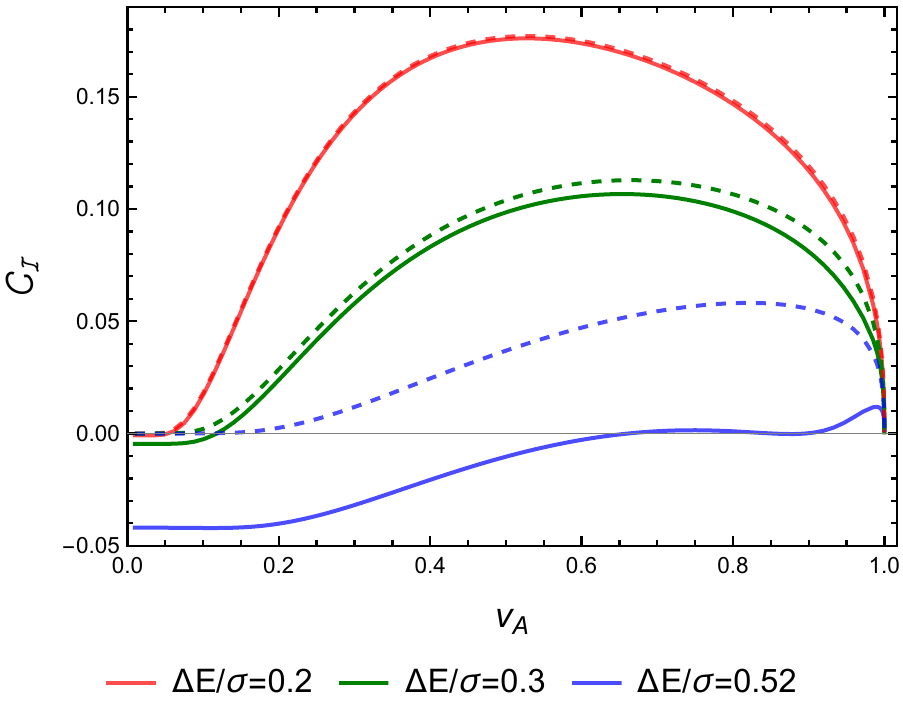}
\caption{In $(1+3)$ dimensions the integral $\mathcal{C}_{\mathcal{I}}$ signifying 
the concurrence is plotted as a function of $v_{A}$ for $v_{B} = 0$. The upper and the lower plots 
respectively correspond to $r_{0}\,\sigma = 0.5$ and $r_{0}\,\sigma = 1$, and these plots are the 
same for both parallel and perpendicular motions (see Eq. (\ref{eq:KogaResult}), and we have also 
checked this by plotting for the parallel and perpendicular cases). In both of these plots we have 
considered the distribution function $f(\omega_{k})=C~\omega_{k} 
~e^{-(\omega_{k}-\omega_{0})^{2}/2\sigma^{2}}$ for the singly excited background field state. It 
should be mentioned that with the exponential decaying distribution function also one gets similar 
plots with the same characteristics. Here the other fixed parameter is $\omega_{0}/\sigma=0.5$. From 
these plots it is evident that with decreasing $\Delta E/\sigma$ and $r_{0}\,\sigma$ the concurrence 
increases. The dashed lines denote the contributions from $\mathcal{C}_{\mathcal{I}}^{vac}$, while 
the solid lines denote $\mathcal{C}_{\mathcal{I}}$.}
\label{fig:C-1p3-perp-vB0}1
\end{figure}
%

The plots from Figs. \ref{fig:C-1p3-perp-vDE-ED} and \ref{fig:C-1p3-perp-vDE-GF} of $\mathcal{C}_{\mathcal{I}}$ with respect to the transition energy $\Delta E$ in this 
perpendicular case represent nature similar to the parallel case (see Fig. 
\ref{fig:C-1p3-para-vDE-ED} and \ref{fig:C-1p3-para-vDE-GF}). From these figures we observe that with increasing detector transition 
energy $\Delta E$ the harvesting decreases (only when $\mathcal{C}_{\mathcal{I}}$ is positive). In Fig. \ref{fig:C-1p3-perp-ED} and \ref{fig:C-1p3-perp-GF} we have plotted 
$\mathcal{C}_{\mathcal{I}}$ as functions of $v_{A}$ for fixed $v_B\neq 0$. 
We have also plotted the quantity $\mathcal{C}_{\mathcal{I}}^{vac} = 
\left|\mathcal{I}_{\varepsilon}\right|$ in these figures, which denotes the concurrence if the 
detectors were interacting with the field vacuum rather than the non-vacuum state. From these plots 
one can observe that entanglement harvesting from the single particle excited state is 
suppressed compared to the vacuum fluctuations. 
Here also, one can observe that in the limit of $v_{A}\to 1$, the quantity 
$\left|\mathcal{I}_{\varepsilon}\right|$ vanishes, leading to a vanishing concurrence, and this 
feature is evident in these figures. When the $\Delta E$ value is large the entanglement harvesting is happening in discrete intermediate and high velocity regimes. 
%
Like the parallel case here also entanglement 
harvesting decreases with increasing distance $r_{0}/\alpha$ between the perpendicular paths, and 
with increasing transition energy $\Delta E$. 
Here when the two detectors are in perpendicular inertial motion in $(1+3)$ dimensions the integral 
$\mathcal{I}_{\varepsilon_{R}}^{vac}$ does not vanish in the $v_{A}\to v_{B}$ limit unlike the 
parallel case. However, its expression in the $v_{B}\to 0$ limit is the same as the parallel case. We also observe that in the $v_{B} = 0$ situation  
(plots corresponding to this situation are given in Fig. \ref{fig:C-1p3-perp-vB0}) the plots are the 
same as the parallel case (done in \cite{Koga:2018the}, where one detector is taken to be 
stationary).  While for $v_{B} \neq 0$ they are much different. 
%

\section{Discussion}\label{sec:discussion}

In the literature, most of the entanglement harvesting-related works, up to our knowledge, consider 
the detectors interacting with the field vacuum. However, in nature, one cannot expect that the 
background field will always be in a vacuum state. Then, the consideration of non-vacuum field 
states in understanding entanglement harvesting becomes essential from the practical point of view. 
The present work has considered single-particle background field excitations to investigate the 
entanglement harvesting between inertial Unruh-DeWitt detectors in $(1+1)$ and $(1+3)$ dimensions. 
Choice of single particle state is motivated from the earlier work \cite{Lochan:2014xja}, and to 
build an analytically simple, tractable model. We have presented a thorough formulation for 
understanding entanglement harvesting from non-vacuum field states.
Our main observation is that in both $(1+1)$ and $(1+3)$ dimensions, the entanglement harvested due 
to the single-particle background fluctuations is reduced compared to the vacuum fluctuations.
The transition probability for inertial detectors interacting with the non-vacuum field state is 
non-zero and non-trivial (see Figs. \ref{fig:Ia-1p1-ED}, \ref{fig:Ia-1p1-GF}, \ref{fig:Ij-1p3-ED}, 
and \ref{fig:Ij-1p3-GF}). This non-zero transition probability is expected even if the detectors are 
inertial as they interact with non-vacuum field states.
We also observed, both in $(1+1)$ and $(1+3)$ dimensions with detectors in parallel motion, that two 
detectors with the same velocity ($v_{A} = v_{B}$) do not harvest any entanglement among themselves 
when interacting with the single particle field state. This phenomenon was true even when the 
detectors interacted with the field vacuum. However, a similar situation with detectors in 
perpendicular inertial motion in $(1+3)$ dimensions is not observed. One observes 
increasing entanglement harvesting with decreasing detector transition energy in both these 
dimensions, for parallel and perpendicular motions, and the considered system parameters. In 
particular, in $(1+1)$ dimensions, there is  visible occurrence of entanglement harvesting in low 
detector transition energy and high-velocity regimes (see Figs. \ref{fig:C-1p1-ED} and 
\ref{fig:C-1p1-GF}).
While in $(1+3)$ dimensions (both with detectors in parallel and perpendicular motion), one notices 
that entanglement harvesting stops when the velocity of detector $A$ approaches the velocity of 
light, i.e., when $v_{A}\to 1$ (see Figs. \ref{fig:C-1p3-para-ED}, \ref{fig:C-1p3-para-GF}, 
\ref{fig:C-1p3-perp-ED}, \ref{fig:C-1p3-perp-GF}, and \ref{fig:C-1p3-perp-vB0}). In $(1+3)$ 
dimensions the entanglement harvesting decreases with increasing perpendicular distance ($r_{0}$) 
between the two detectors. 

We want to provide a few final comments, which are as follows. First, we emphasize that entanglement harvesting is possible only when $|\mathcal{I}_{\epsilon}|^2>\mathcal{I}_{A}\mathcal{I}_{B}$. The nonnegative measure of $\mathcal{C}_{\mathcal{I}} = |\mathcal{I}_{\epsilon}| - \sqrt{\mathcal{I}_{A}\mathcal{I}_{B}}$ quantifies the harvested entanglement. Interestingly the local terms $\mathcal{I}_{j}$ (with $j$ being either $A$ or $B$) denote individual detector transition probabilities. These terms have a vanishing contribution to inertial detectors interacting with the vacuum background scalar field. While for the background field in any excited state, these local terms for the similar detector motion are finite and nonzero. Therefore, the harvesting from the nonvacuum states is always expected to be lesser compared to the field vacuum if the entangling term $\mathcal{I}_{\epsilon}$ does not change much. In our current work, we found that this is indeed the case with single-particle excited states, as the entangling term $\mathcal{I}_{\epsilon}$ remains the same for the vacuum and nonvacuum field states.

Furthermore, we found some additional distinct features with the single-particle excited states compared to the vacuum field state. For example, for two inertial detectors moving at different velocities, entanglement harvesting from the field vacuum always seems possible as one varies the detector transition energy $\Delta E$. While the same is not true from the single-particle excited state as one can perceive ranges of no-harvesting in $\Delta E$. One observes this phenomenon in both $(1+1)$ and $(1+3)$ dimensions and for parallel and perpendicular cases. In this regard, see the plots of concurrence as a function of $\Delta E$ in Figs. \ref{fig:C-1p1-ED}, \ref{fig:C-1p1-GF}, \ref{fig:C-1p3-para-vDE-ED}, \ref{fig:C-1p3-para-vDE-GF}, \ref{fig:C-1p3-perp-vDE-ED}, and \ref{fig:C-1p3-perp-vDE-GF}. When examining the results with a varying velocity $v_{A}$ of the detector $A$, one may point out a similarly interesting distinguishing feature in the perpendicular motion case. In this scenario, one can observe discrete regions of entanglement harvesting in $v_{A}$ from the single-particle field states for larger detector transition energies. In comparison, in similar scenarios, entanglement harvesting from the vacuum is possible in the whole range $0<v_{A}<1$, see Figs. \ref{fig:C-1p3-perp-ED} and \ref{fig:C-1p3-perp-GF}. These specific observations provide additional distinctions between the vacuum and single-particle entanglement harvesting cases other than the expected diminishing phenomenon observed for the latter case. Finally, one should notice that in nature, it is not guaranteed that the background field state will be in a vacuum. In fact, it is natural to believe that the fields will be in some excited state due to the presence of various constituents. Then, these specific features of harvesting shall provide some valuable distinctions of identification. Moreover, for the physical realization of entanglement harvesting from the single-particle field state, there is a restriction in the range of system parameters (prominently visible in $\Delta E$), unlike the vacuum. Therefore, contrary to the vacuum case, to observe the entanglement harvesting, one needs to construct a detector such that the transition energy must live within the allowed range of values.

Secondly, we shall like to emphasize that the nature of the individual detector transition probabilities $\mathcal{I}_{j}$ alters considerably depending on the motion of the detectors. For instance, inertial detectors have vanishing transition probabilities for interactions with the field vacuum. In contrast, the non-inertial detectors have non-vanishing contributions, which results in the Unruh effect for uniform linear acceleration. Therefore, the scenarios with non-inertial detectors demand an extensive and detailed investigation, and it is difficult to comment on these scenarios just by studying the inertial case. Nevertheless, entanglement harvesting from the non-vacuum field states with non-inertial detectors (see Refs. \cite{Koga:2019fqh, Barman:2021oum, Barman:2021bbw}) remains an interesting arena to venture further. In this regard, we believe the cases of charged detectors in circular trajectories \cite{Bell:1982qr, Bell:1986ir, Akhmedov:2006nd, Akhmedov:2007xu} compels great enthusiasm as these systems open the avenue to realize many theoretical predictions physically.

We believe this work opens up a new avenue to understanding the nature of entanglement harvesting from non-vacuum field states. The immediate direction to look further, discussed in the previous paragraph and which we are currently working on, is entanglement harvesting from these non-vacuum field states with non-inertial detectors. In particular, some of our preliminary observations, corresponding to the uniformly accelerated linear motion of the detectors, are as follows. The excess contribution due to the non-vacuum fluctuations in the individual detector transition probability $\mathcal{I}_{j}$ is finite for detectors with infinite switching. Thus the relevant transition rate, coming from this contribution, over infinite time vanishes. In comparison, one observes that the same contribution $\mathcal{I}_{j}$ due to the vacuum fluctuation has a Dirac delta zero multiplying factor (see the transition probabilities of non-inertial Unruh-DeWitt detectors from \cite{Koga:2019fqh, Barman:2021oum, Barman:2021bbw}), which results in a finite transition rate for infinite time. Therefore, in concurrence per unit time, the different effects of vacuum and non-vacuum fluctuations will not be evident. We have considered detectors with finite time switching functions to circumvent this inconvenience, which provides finite transition probabilities. In future work, we wish to communicate the progress in this direction and with detectors in a circular motion. Our endeavor also includes investigating and comparing entangled atoms' transition probability characteristics with the so-called Unruh effect of the accelerated observers for detectors interacting with a non-vacuum background field state.

\begin{acknowledgments}
DB would like to acknowledge Ministry of Education, Government of India for
providing financial support for his research via the PMRF May 2021 scheme. SB would like to thank the Indian 
Institute of Technology Guwahati (IIT Guwahati) for financial support. The research of BRM is 
partially supported by a START-UP RESEARCH GRANT (No. SG/PHY/P/BRM/01) from the Indian Institute of 
Technology Guwahati, India.

\end{acknowledgments}

\begin{appendix}

\section{Normalized distribution functions for singly excited states}\label{Appn:dist-fns}
\subsection{In $(1+1)$ dimensions}\label{Appn:dist-fns-1p1}

One may consider qualitatively different distribution functions $f(\omega_{k})$ (see 
\cite{Lochan:2014xja}). In $(1+1)$ dimensions a few examples follow, and we also provide their 
respective normalization constants. It should be noted that to evaluate the normalization constant 
in $f(\omega_{k})$, here one has to use the $(1+1)$ dimensional form of the normalization condition 
(\ref{eq:Normalization-cond-fw}).

\subsubsection{Exponential damping, $f(\omega_{k}) = C\,\omega_{k}\,e^{-\alpha\omega_{k}}$} 
In this case the distribution function is of the form 
$f(\omega_{k}) = C\,\omega_{k}\,e^{-\alpha\omega_{k}}$. Using the normalization condition 
(\ref{eq:Normalization-cond-fw}) of $f(\omega_{k})$ for $(1+1)$ dimensions one can find out the 
value of 
the constant $C$ as $C=2 \,\sqrt{2 \pi }\,\alpha$.

\subsubsection{Gaussian distribution, $f(\omega_{k}) = C\,\omega_{k} 
\,e^{-(\omega_{k}-\omega_{0})^{2}/2\sigma^{2}}$} 
In this case the distribution function is of the form $f(\omega_{k}) = C\,\omega_{k} 
\,e^{-(\omega_{k}-\omega_{0})^{2}/2\sigma^{2}}$. Using the normalization condition 
(\ref{eq:Normalization-cond-fw}) of $f(\omega_{k})$ for $(1+1)$ dimensions one can find out the 
value of 
the constant $C$ here, as $C=\left\{4\sqrt{\pi}/ 
\left[\omega_{0}\,\sigma\left(2-Q\left(-1/2,\omega_{0}^2/\sigma ^2\right) \right) \right] 
\right\}^{1/2}$. Here $Q\left(a,z\right)$ denotes the Regularized Gamma Function.

\subsection{In $(1+3)$ dimensions}\label{Appn:dist-fns-1p3}
In $(1+3)$ dimensions also one may consider similar distribution functions 
$f(\omega_{k})$. However, here one has to use the $(1+3)$ dimensional form of the normalization 
condition (\ref{eq:Normalization-cond-fw}) to find out the respective normalization 
constants. Naturally the normalization constants $C$ obtained here will be different than the ones 
obtained for $(1+1)$ dimensional case.

\subsubsection{Exponential damping, $f(\omega_{k}) = C\,\omega_{k}\,e^{-\alpha\omega_{k}}$} 
Like the $(1+1)$ dimensional case here also we consider the same exponentially damping distribution 
function
\begin{equation}\label{f1}
f(\omega_{k})=C~\omega_{k}~e^{-\alpha\omega_{k}}~.
\end{equation}
However, here the normalization constant $C$ is obtained from Eq. (\ref{eq:Normalization-cond-fw}) 
by putting $d=3$, which corresponds to three spatial dimensions of our present case. Then this 
constant $C$ is given by $C=4 \pi\alpha ^2\,\sqrt{2/3}~$.

\subsubsection{Gaussian distribution, $f(\omega_{k}) = C\,\omega_{k} 
\,e^{-(\omega_{k}-\omega_{0})^{2}/2\sigma^{2}}$} 
We consider the Gaussian distribution function for a singly excited state in $(1+3)$ dimensions as
\begin{equation}\label{f2}
f(\omega_{k})=C~\omega_{k} ~e^{-(\omega_{k}-\omega_{0})^{2}/2\sigma^{2}}.
\end{equation}
Here also one has to use the normalization condition (\ref{eq:Normalization-cond-fw}) 
with $d=3$ to evaluate the value of the normalization constant $C$, which is given by 
\begin{eqnarray}
C&=&4\pi\Big/\bigg\{\sqrt{\pi } \omega_{0} \left(3 \sigma^2+2 \omega_{0}^2\right)\sigma 
\left(\textup{Erf}\left(\frac{\omega_{0}}{\sigma 
}\right)+1\right)\no\\&&~~~~~~~~~~~~~~+2 
\sigma^{2}  e^{-\frac{\omega_{0}^2}{\sigma^2}} \left(\sigma^2+\omega_{0}^2\right)\bigg\}^{1/2}~,
\end{eqnarray}
where, $\textup{Erf}(z)$ denotes the error function.

\section{Evaluations of the integrals in $(1+1)$ dimensions}\label{Appn:1p1-integrals}
In this section of the Appendix we provide an explicit derivation of the integral 
$\mathcal{I}_{\varepsilon_{R}}^{vac}$. In particular, we shall be obtaining its expression as 
provided in Eq. (\ref{eq:IeRvac-1p1-1}) for two inertial detectors in $(1+1)$ dimensions. In this 
regard we consider the form of the integral from Eq. (\ref{eq:Ie-integral-b}) with the consideration 
that $T_{j}=t_{j}$ are now the Minkowski times. Then this integral becomes
\begin{eqnarray}\label{eq:}
 \mathcal{I}_{\varepsilon_{R}}^{vac} 
&=& \frac{1}{\gamma_{A}\gamma_{B}} \intinf 
\frac{dk}{4\pi\omega_{k}}\,\int_{-\infty}^{\infty}dt_{A} 
\int_{-\infty}^{\infty}dt_{B}\,\theta(t_{A}-t_{B})\no\\
~&\times& e^{i\Delta E(t_{B}/\gamma_{B}+t_{A}/\gamma_{A})} \,\Big[ e^{
-i(\omega_{k}-k\,v_{B})\,t_{B}
+i(\omega_{k}-k\,v_{A})\,t_{A}}\no\\
~&& ~~ - e^{
i(\omega_{k}-k\,v_{B})\,t_{B}
-i(\omega_{k}-k\,v_{A})\,t_{A}}\Big]\no\\
&=& \frac{1}{\gamma_{A}\gamma_{B}} \intinf 
\frac{dk}{4\pi\omega_{k}}\,\int_{-\infty}^{\infty}dt_{A} 
\int_{-\infty}^{t_{A}}dt_{B}\no\\
~&\times& e^{i\Delta E(t_{B}/\gamma_{B}+t_{A}/\gamma_{A})} \,\Big[ e^{
-i(\omega_{k}-k\,v_{B})\,t_{B}
+i(\omega_{k}-k\,v_{A})\,t_{A}}\no\\
~&& ~~ - e^{
i(\omega_{k}-k\,v_{B})\,t_{B}
-i(\omega_{k}-k\,v_{A})\,t_{A}}\Big]\no\\
&=& \frac{i}{\gamma_{A}\gamma_{B}}\intinf 
\frac{dk}{4\pi\omega_{k}} 
\,\int_{-\infty}^{\infty}dt_{A}\,e^{i\Delta 
E\,t_{A}(1/\gamma_{B}+1/\gamma_{A})}\,\no\\
~&\times& \Bigg[\frac{e^{-
i\{k\,(v_{A}-v_{B})+i\,\epsilon\}\,t_{A}
}}{\omega_{k}-k 
\,v_{B}-\Delta{E}/\gamma_{B}}  + \frac{e^{
i\,k\{(v_{A}-v_{B})-i\,\epsilon\}\,t_{A}
}}{\omega_{k}-k 
\,v_{B}+\Delta{E}/\gamma_{B}}\Bigg]\,,\no\\
\end{eqnarray}
where one can observe that a $e^{\epsilon\,t_{B}}$ regulator, with positive small $\epsilon$, was 
introduced to make the $t_{B}$ integral convergent in the lower limit of negative infinity. The 
actual integral is obtained in the limit of $\epsilon\to 0$. After performing the $t_{A}$ 
integration now one can obtain 
\begin{eqnarray}\label{eq:}
 \mathcal{I}_{\varepsilon_{R}}^{vac}
&=& \frac{i}{\gamma_{A}\gamma_{B}}\intinf 
\frac{dk}{2\omega_{k}} 
\,\Bigg[\tfrac{\delta\left[k(v_{A}-v_{B})-\Delta{E}\left(\frac{1}{\gamma_{A}}+\frac{1}{\gamma_{B}}
\right) \right] } { \omega_{k}-k 
\,v_{B}-\Delta{E}/\gamma_{B}} \no\\
~&& +~ 
\tfrac{\delta\left[k(v_{A}-v_{B})+\Delta{E}\left(\frac{1}{\gamma_{A}}+\frac{1}{\gamma_{B}}
\right)\right ] } { \omega_{k}-k 
\,v_{B}+\Delta{E}/\gamma_{B}}\Bigg] \,.
\end{eqnarray}
This expression is the same as the one presented in Eq. (\ref{eq:IeRvac-1p1-1}).

\section{Evaluations of the integrals in $(1+3)$ dimensions for 
detectors in parallel motion}\label{Appn:1p3-parallel-integrals}

\subsection{Explicit evaluation of $\mathcal{I}_{A}$}\label{Appn:1p3-parallel-integrals-IA}
Here we consider evaluating the necessary integrals for entanglement harvesting for two detectors 
in parallel inertial motion in $(1+3)$ dimensions. To derive the expression of the quantity 
$\mathcal{I}_{A}$ corresponding to detector $A$ we consider the trajectory from 
Eq. (\ref{eq:ParaTrajecA}) and the $\mathcal{A}(\Delta E)$ part of it becomes
\begin{eqnarray}\label{1p3Pj}
\mathcal{A}(\Delta E) &=&\intinf d\tau_{A} e^{i\Delta 
E\tau_{A}}\Phi_{\mathrm{eff}}(x)\no\\&=&\intinf d\tau_{A} e^{i\Delta 
E\tau_{A}}\int_{0}^{\pi}\intsinf \frac{2\pi\sin\theta\omega_{k}^{2}d\omega_{k} d\theta}{(2 
\pi)^{3}} 
\no\\&&~~~~~~~~~~~~~~~~\times\frac{f(\omega_{k})}{2 \omega_{k}}~  
e^{-i\omega_{k}\gamma_{A}\tau_{A}+i\omega_{k}\gamma_{A} v\tau_{A}\cos\theta}\no\\
&=&-\frac{i}{2(2 
\pi)^{2}\gamma_{A} 
v_{A}}\intsinf d\omega_{k}f(\omega_{k}) \intinf 
\frac{d\tau_{A}}{\tau_{A}}\no\\&&~~ \times~ [e^{i \tau_{A}(\Delta 
E-\omega_{k}/D_{A})}-e^{i \tau_{A}(\Delta 
E-\omega_{k} 
D_{A})}]\,,
\end{eqnarray}
where $D_{j} = \sqrt{(1+v_{j})/(1-v_{j})}$ and $\theta$-integration is done to obtain the final 
result. Now it is known that 
\begin{eqnarray}
\intinf d\tau \frac{ e^{i \tau\alpha}}{\tau}\no&=&\intinf d\tau \frac{ \cos{ 
\tau\alpha}}{\tau}+i\intinf d\tau \frac{ \sin{ \tau\alpha}}{\tau}\no\\&=&i\pi\,\text{sgn}(\alpha)
\end{eqnarray}
where $\operatorname{sgn}(\alpha)$ is $\pm1$ depending on positive or negative sign of the $\alpha$, 
or $zero$  for $\alpha=0$. Thus the expression of (\ref{1p3Pj}) is non-zero only when, 
$1/D_{A}\leq\omega_{k}/\Delta E\leq D_{A}$, i.e., when $(\Delta E-\omega_{k}/D_{A})>0 $ with 
$(\Delta E-\omega_{k} D_{A})<0$. Therefore (\ref{1p3Pj}) becomes
\begin{equation}\label{IAgen}
\mathcal{A}(\Delta E) = \frac{1}{4 \pi\gamma_{A} v_{A}}\int_{\Delta 
E/D_{A}}^{D_{A}\Delta E} d\omega_{k} f(\omega_{k})~.
\end{equation}
Similarly one can check that
\begin{eqnarray}
\mathcal{B}(\Delta E) &=&
\intinf d\tau_{A}\, e^{i\Delta E\,\tau_{A}}\Phi^{\star}_{\mathrm{eff}}(x)\no\\
~&=&-\frac{i}{2(2 \pi)^{2}\gamma_{A} v_{A}}\int 
d\omega_{k} f(\omega_{k})[i\pi-i\pi]\no\\
~&=& 0~.
\end{eqnarray}
%

\subsection{Explicit evaluation of $\mathcal{I}_{B}$}\label{Appn:1p3-parallel-integrals-IB}
The quantity $\mathcal{A}(\Delta E)$ for detector $B$ from Eq. (\ref{eq:Ijv-Ijnv-2a}) can be 
evaluated as
\begin{eqnarray}\label{1p3Pj0}
&&\mathcal{A}(\Delta E)\no\\&&=\intinf d\tau_{B} e^{i\Delta 
E\tau_{B}}\Phi_{\mathrm{eff}}(x_{B})\no\\&&=\intinf 
d\tau_{B} 
e^{i\Delta E\tau_{B}}\int_{0}^{2\pi}\int_{0}^{\pi}\intsinf 
\frac{d\phi\sin\theta\omega_{k}^{2}d\omega_{k} d\theta}{(2 \pi)^{3}} \frac{f(\omega_{k})}{2 
\omega_{k}} \no\\&&~~~~\times 
e^{-i\omega_{k}\gamma_{B}\tau_{B}+i\omega_{k}\sin\theta(x_{0}\cos\phi+y_{0}\sin\phi)+i\omega_{k}
v_{B}\gamma_{B}\tau_{B}\cos\theta}\no\\&&=
\intinf d\tau_{B} e^{i\Delta E\tau_{B}}\int_{0}^{\pi}\intsinf  \frac{\sin\theta d\theta\omega_{k} 
d\omega_{k}}{(2 \pi)^{2}} \frac{f(\omega_{k})}{2 } \no\\&&~~~~~~~~~~~~~~~\times 
e^{-i\omega_{k}\gamma_{B}\tau_{B}(1-v_{B}\cos\theta)} J_{0}(
\omega_{k}r_{0}\sin\theta)\no\\&&=
\int_{-1}^{1}du \intsinf \frac{\omega_{k} d\omega_{k}}{(2 \pi)} \frac{f(\omega_{k})}{2 } 
\delta(\Delta E-\omega_{k}\gamma_{B}(1-v_{B}u))\no\\&&~~~~~~~~~\times  
J_{0}(\omega_{k}r_{0}\sqrt{1-u^{2}})\no\\&&=
\int_{-1}^{1}\frac{du}{4\pi} \frac{\Delta E~f\left(\frac{\Delta 
E}{\gamma_{B}(1-v_{B}u)}\right)}{\{\gamma_{B}(1-v_{B}u)\}^{2}}J_{0}\left(
\tfrac{r_{0}\Delta E\sqrt{1-u^{2}}}{\gamma_{B}(1-v_{B}u)}\right),
\end{eqnarray}
where after the third equality we have defined $u=\cos\theta$. We can evaluate the integration over 
$u$ in (\ref{1p3Pj0}) numerically using the expressions of $f(\omega_{k})$ given in (\ref{f1}) and 
(\ref{f2}), with the appropriate forms of the normalization constant $C$ separately in $(1+1)$ and 
$(1+3)$ dimensions.

\subsection{Explicit evaluation of $\mathcal{I}_{\varepsilon}$}

\subsubsection{Considering the expression of the Green's function in position 
space}\label{Appn:1p3-parallel-integrals-Ie-position}

Here we consider the position space representation of the Green's function appearing in the 
expression of the integral
\begin{equation}\label{eq:IeVacR-Appn}
\mathcal{I}_{\varepsilon_{R}}^{vac}=-i\int_{-\infty}^{\infty} d \tau_{B} \int_{-\infty}^{\infty} d 
\tau_{A}  e^{i \Delta E( \tau_{A}+\tau_{B})} G_{R}\left(X_{A}, X_{B}\right).
\end{equation}
The retarded Green function $G_{R}\left(X_{A}, X_{B}\right)$ corresponding to a massless, minimally 
coupled free scalar field in the Minkowski spacetime is
\begin{eqnarray}
G_{R}\left(X_{A},X_{B}\right) &=& -
\frac{1}{2 \pi}~\Theta
\left(t_{A}-t_{B}\right) \no\\
~&& \times\,\delta
\left(\left(t_{A}-t_{B}\right)^{2}-\left|\boldsymbol{x}_{A}-\boldsymbol{x}_{B}
\right|^{2}\right)
\no\\&=& -\frac{1}{2 \pi} \Theta\left(t_{A}-t_{B}\right) 
\delta(g(t_{A},t_{B}))~.
\end{eqnarray}
For two detectors moving in parallel inertial trajectories (see Eq. (\ref{eq:ParaTrajecA}) and 
(\ref{eq:ParaTrajecB})) in $(1+3)$ dimensions the argument of the Dirac delta distribution becomes 
zero when
\begin{eqnarray}\label{gttp}
g(t_{A},t_{B}) &=& \scalebox{0.85}{${A}^{2}(1-v_{A}^{2})+t_{B}^{2}(1-v_{B}^{2})-2 t_{A} 
t_{B}(1-v_{A}v_{B})-r_{0}^{2}$}\no\\
~&=& 0~.
\end{eqnarray}
This equation has solutions for $t_{B}$ as
\begin{equation}\label{MinGDeno}
t_{B}=\gamma_{B}^{2}(t_{A}(1-v_{A}v_{B})\pm u(t_{A}))\equiv t_{\pm}
\end{equation}
where, 
\begin{equation}\label{utA}
u(t_{A})=\sqrt{t_{A}^{2}(v_{A}-v_{B})^{2}+r_{0}^{2}(1-v_{B}^{2})}.
\end{equation}
Now, with the general condition $0\le v_{j}\le 1$, one can check that
\begin{eqnarray}
&&t_{+}>t_{A}~,\no\\&&t_{-}<t_{A}~.
\end{eqnarray}
Then using the property of Dirac delta functions, we have
\begin{eqnarray}
\delta(g(t_{A},t_{B}))&=&\left[\frac{\delta(t_{B}-t_{+})}{|g^{\p}(t_{A},t_{B})|_{t_{B}=t_{+}}}+\frac
{\delta(t_{B}-t_{-})}{|g^{\p}(t_{A},t_{B})|_{t_{B}=t_{-}}}\right]\no\\&=&\frac{1}{2u(t_{A})}[
\delta(t_{B}-t_{+})+\delta(t_{B}-t_{-})]~~
\end{eqnarray}
Therefore one may express the integral of Eq. (\ref{eq:IeVacR-Appn}) as
\begin{eqnarray}\label{1p3perp0}
\mathcal{I}_{\varepsilon_{R}}^{vac} &=& \frac{-i}{2\pi}\intinf
{\frac{dt_{A}}{\gamma_{A}}}\int_{-\infty}^{t_{A}}{\frac{dt_{B}}{\gamma_{B}}} e^{
i\Delta{E}\big(\frac{t_{A}}{\gamma_{A}}+\frac{t_{B}}{\gamma_{B}}\big)}\delta(g(t_{A},t_{B}))\no\\
&=& \frac{-i}{4\pi}\intinf{\frac{dt_{A}}{\gamma_{A}\gamma_{B} 
~u(t_{A})}}e^{i\Delta{E}\left(\frac{t_{A}}{\gamma_{A}}+\gamma_{B}(t_{A}(1-v_{A}v_{B})-u(t_{A}
))\right)}\no\\
&=& \frac{-i}{2\pi 
}\intsinf\scalebox{.95}{$\frac{dt_{A}e^{-i\Delta{E}\gamma_{B}u(t_{A})}}{\gamma_{A}\gamma_{B}\,u(t_{A
})}\cos$}\scalebox{.85}{$\left(\Delta{E}t_{A}\left(\frac{1}{\gamma_{A}}+\gamma_{B}(1-v_{A}v_{B}
)\right)\right)$}~.\no\\
\end{eqnarray}
Now expanding the exponential in (\ref{1p3perp0}) in terms of $sin$ and $cos$ functions and by 
comparing with (\ref{eq:BesselK0-IntRep}), we can recognise 
\begin{eqnarray}
a&=&\frac{r_{0}}{\gamma_{B}|v_{A}-v_{B}|};~~ 
\varrho=\gamma_{B}|v_{A}-v_{B}|\Delta{E};\no\\ 
\beta&=&\Delta{E}\left(\frac{1}{\gamma_{A}}+\gamma_{B}(1-v_{ A } v_
{B})\right);
\end{eqnarray}
as we know $0\le v_{j}\le 1$, this satisfies the criteria for (\ref{eq:BesselK0-IntRep}), $i.e.$,
\begin{eqnarray}
&&\beta>\varrho~\text{or,~}\left(\frac{1}{\gamma_{A}}+\gamma_{B}(1-v_{A}v_{B})\right)>\gamma_{B}|v_{
A } -v_ { B }
|\no\\&&\text{or,~~}\frac{1}{\gamma_{A}(1-v_{A}v_{B})}+\gamma_{B}>\gamma_{B}|v_{rel}|.
\end{eqnarray}
Thus $\beta>\varrho>0$ is satisfies as the relative velocity ($v_{rel}$) between the detectors is 
always less 
than $one$. Therefore, from  (\ref{1p3perp0}), we obtain
\begin{eqnarray}\label{eq:IeVacR-prl-1p3D-Appn}
\mathcal{I}_{\varepsilon_{R}}^{vac} 
&=& \frac{-i}{2\pi\gamma_{A}\gamma_{B}|v_{A}-v_{B}|}\no\\&&\times 
K_{0}\left(\tfrac{r_{0}\Delta{E}}{\gamma_{B}}\sqrt{\left(\tfrac{\frac{1}{\gamma_{A}}+\gamma_{B}(1-v_ 
{ A }v_{B})}{v_{A}-v_{B}}\right)^{2}-\gamma_{B}^{2}}\right)~.\no\\
\end{eqnarray}
If $v_{A}=v_{B}$ then from (\ref{MinGDeno}) and (\ref{utA}), we obtain $u(t_{A})$ is $t_{A}$ 
independent and $t_{-}=t_{A}-r_{0}\gamma_{B}$. Therefore, the $t_{A}$-integration after the second 
last line of (\ref{1p3perp0}) gives a delta function of form $\delta\left(\Delta{E} 
\left(1/\gamma_{A} + 1/\gamma_{B} \right)\right)$. This is always $zero$ for $v_{A}=v_{B}$, as both 
$\gamma_{A}$ and $\Delta{E}$ are always positive. One can also simply take the limit $(v_{A}- 
v_{B})\to 0$ and observe that this quantity readily vanishes from Eq. 
(\ref{eq:IeVacR-prl-1p3D-Appn}).

\subsubsection{Considering the expression of the Green's function in momentum 
space}\label{Appn:1p3-parallel-integrals-Ie-momentum}

On the other hand, considering the expression of the Green's function in momentum space on can 
express the integral $\mathcal{I}_{\varepsilon_{R}}^{vac}$ from Eq. (\ref{eq:IeVacR-Appn}) as
\begin{eqnarray}\label{1p3parallel}
\mathcal{I}_{\varepsilon_{R}}^{vac} &=&-\intinf d\tau_{A}\intinf d\tau_{B} 
~e^{i\Delta{E}(\tau_{A}+\tau_{B})}\theta(\gamma_{A}\tau_{A}-\gamma_{B}\tau_{B}
)\no\\&&~~~\times\int_{0}^{\pi}\sin\theta d\theta \intsinf\frac{\omega_{k}d\omega_{k}}{2(2 
\pi)^{2}} 
\Big[e^{-i\omega_{k}(\gamma_{A}\tau_{A}-\gamma_{B}\tau_{B})}\no\\&&~~~~~~~~\times 
e^{i\omega_{k}\sqrt{(\gamma_{A}\tau_{A} v_{A}-\gamma_{B}\tau_{B}v_{B})^{2}+ 
r_{0}^{2}}\cos\theta}\no\\&&-e^{i 
\omega_{k}(\gamma_{A}\tau_{A}-\gamma_{B}\tau_{B})-i\omega_{k}\sqrt{(\gamma_{A}\tau_{A} 
v_{A}-\gamma_{B}\tau_{B}v_{B})^{2}+r_{0}^{2}}\cos\theta }\Big]
\no\\
&=&	-\intinf \intinf \frac{dp~dq}{2v_{A}v_{B}'\gamma_{A}\gamma_{B}} 
e^{i\Delta{E}(a_{1}p+a_{2}q)}\theta(a_{3}p-q)\no\\&&\times\int_{-1}^{1}du\intsinf 
\frac{\omega_{k}d\omega_{k}}{2(2 \pi)^{2} } 
\Big[e^{-i\omega_{k}a_{4}(a_{3}p-q)+i\omega_{k}\sqrt{p^{2}+ 
r_{0}^{2}}u}\no\\&&~~~~~~~~~~~~-e^{i 
\omega_{k}a_{4}(a_{3}p-q)-i\omega_{k}\sqrt{p^{2}+r_{0}^{2}}u }\Big]\,.
\end{eqnarray}
As we have already discussed in sec. \ref{subsec:1p3D-parallel-UV}, here we defined a change of 
variables, $p=v_{A}\gamma_{A}\tau_{A}-v_{B}\gamma_{B}\tau_{B}$ and 
$q=v_{A}\gamma_{A}\tau_{A}+v_{B}\gamma_{B}\tau_{B}$. The Jacobian of the transform is given by 
$|J|=1/(2v_{A}v_{B}\gamma_{A}\gamma_{B})$. Under this transform the quantities 
$(\gamma_{A}\tau_{A}-\gamma_{B}\tau_{B})$ and $(\tau_{A}+\tau_{B})$, can be expressed as
\begin{eqnarray}
&&\gamma_{A}\tau_{A}-\gamma_{B}\tau_{B}=\frac{p+q}{2v_{A}}-\frac{q-p}{2v_{B}}=\scalebox{1.1}{$\frac{
p(v_{B}+v_{A})+q(v_{B}-v_{A})}{2v_{A}v_{B}}$}\no\\&&~~~~~~=\frac{v_{A}-v_{B}}{2v_{A}v_{B}}
\left(p\frac{v_{A}+v_{B}}{v_{A}-v_{B}}-q\right)=a_{4}(a_{3}p-q);\no\\
&&\tau_{A}+\tau_{B}=p\scalebox{1.1}{$\left(\frac{1}{2v_{A}\gamma_{A}}-\frac{1}{2v_{B}\gamma_{B}}
\right)$}+q\scalebox{1.1}{$\left(\frac{1}{2v_{A}\gamma_{A}}+\frac{1}{2v_{B}\gamma_{B}}\right)$}
\no\\&&~~~~~~~~=a_{1}p+a_{2}q;\label{CoordTranspq}
\end{eqnarray}
In Eq. (\ref{1p3parallel}) the step function provides the upper limit of 
$q$-integration, which is $a_{3}p$. After evaluating the $q$-integral, we obtain
\begin{eqnarray}
 \mathcal{I}_{\varepsilon_{R}}^{vac} &=&-\intinf  
\frac{dp~e^{i\Delta{E}a_{1}p}}{2vv'\gamma'\gamma}\int_{-1}^{1}du\intsinf 
\frac{\omega_{k}d\omega_{k}}{8\pi^{2}} \no\\
~&&\times\,\Big[ \frac{e^{i\Delta{E}a_{2}a_{3}p}}{i(\Delta{E}a_{2}+a_{4}\omega_{k})} \, 
e^{i\omega_{k}\sqrt{p^{2}+ 
r_{0}^{2}}u}\no\\
~&&-\frac{e^{i\Delta{E}a_{2}a_{3}p}}{i(\Delta{E}a_{2}-a_{4}\omega_{k})} 
e^{-i\omega_{k}\sqrt{p^{2}+r_{0}^{2}}u }\Big]~.
\end{eqnarray}
Then we perform the $u$-integral and after some re-arrangement we obtain
\begin{eqnarray}\label{1p3parallel1}
\mathcal{I}_{\varepsilon_{R}}^{vac} &=& \intinf  
\frac{dp~e^{i\Delta{E}p(a_{1}+a_{2}a_{3})}}{16\pi^{2}vv'\gamma'\gamma\sqrt{p^{2}+r_{0}^{2}}} 
\Bigg[\intinf \frac{d\omega_{k}~e^{i\omega_{k}\sqrt{p^{2}+ 
r_{0}^{2}}}}{a_{4}\omega_{k}+a_{2}\Delta{E}}\no\\&&~~~~~~~~~~+\intinf 
\frac{d\omega_{k}~e^{i\omega_{k}\sqrt{p^{2}+ r_{0}^{2}}}}{a_{4}\omega_{k}-a_{2}\Delta{E}} \Bigg].
\end{eqnarray}
To evaluate the $\omega_{k}$-integration we change the integration variable to $a_{4}\omega_{k}\pm~ 
a_{2}\Delta{E}$ for first and second integration, respectively. The integration limit remain 
unchanged as the quantities $a_{4},~a_{2}\Delta{E}$ are finite. Then we will use an identity from 
complex variable theory, that is
\begin{eqnarray}
\intinf{dx}\frac{e^{i\alpha x}}{x}=i\pi~\text{sgn}(\alpha)~,
\end{eqnarray}
then the expression from (\ref{1p3parallel1}) will become 
\begin{eqnarray}\label{1p3parallel2}
\mathcal{I}_{\varepsilon_{R}}^{vac} &=& i\intinf  \tfrac{dp~e^{i\Delta{E}p(a_{1}+a_{2}a_{3})}}{8\pi 
v_{A}v_{B}\gamma_{A}\gamma_{B} 
a_{4}\sqrt{p^{2}+r_{0}^{2}}} ~\cos\left(\tfrac{a_{2}\Delta{E}\sqrt{p^{2}+ 
r_{0}^{2}}}{a_{4}}\right)\no\\&&=i\intsinf  
\tfrac{dp~\cos\left(\frac{a_{2}\Delta{E}}{a_{4}}\sqrt{p^{2}+ 
r_{0}^{2}}\right)}{2\pi (v_{A}-v_{B})\gamma_{A}\gamma_{B}\sqrt{p^{2}+r_{0}^{2}}} 
\cos\left(\Delta{E}p(a_{1}+a_{2}a_{3})\right)\no\\&&=\tfrac{i~K_{0}\left(r_{0}\Delta{E}\sqrt{
\left(a_{1}+a_{2}a_{3}\right)^{
2
}-\left(a_{2}/a_{4}\right)^{2}}\right)}{2\pi (v_{A}-v_{B})\gamma_{A}\gamma_{B}}~,
\end{eqnarray}
where we have used the integral representation from Eq. (\ref{eq:BesselK0-IntRep}).

In a manner similar to the previous case one can also express the integral  
$\mathcal{I}_{\varepsilon_{W}}^{vac}$ as
\begin{eqnarray}\label{1p3parallelW}
\mathcal{I}_{\varepsilon_{W}}^{vac} &=& -\intinf d\tau_{A}\intinf d\tau_{B} 
e^{i\Delta{E}(\tau_{A}+\tau_{B})}G_{W}(X_{A},X_{B})\no\\
&=&-\intinf d\tau_{A}\intinf d\tau_{B} e^{i\Delta{E}(\tau_{A}+\tau_{B})} 
\int_{0}^{\pi}\sin\theta \,d\theta\no\\
~&& \times\intsinf \frac{\omega_{k}d\omega_{k}}{2(2 \pi)^{2}}\,
e^{-i\omega_{k}(\gamma_{A}\tau_{A}-\gamma_{B}\tau_{B})}\no\\
~&& \times~e^{i\omega_{k}\sqrt{(\gamma_{A}\tau_{A} 
v_{A}-\gamma_{B}\tau_{B}v_{B})^{2}+ r_{0}^{2}}\cos\theta}~.
\end{eqnarray}
%
%
Using the previously mentioned coordinate transformation, related to Eq. (\ref{CoordTranspq}), the 
quantity from Eq. (\ref{1p3parallelW}) can be evaluated as
\begin{eqnarray}\label{1p3parallelW1}
\mathcal{I}_{\varepsilon_{W}}^{vac} &=&-\intinf \intinf 
\frac{dp~dq~e^{i\Delta{E}(a_{1}p+a_{2}q)}}{2v_{A}v_{B}\gamma_{A}\gamma_{B}} 
\int_{0}^{\pi}\sin\theta d\theta\no\\
~&& \times ~\intsinf \frac{\omega_{k}d\omega_{k}}{2(2 \pi)^{2}}\, 
 e^{-i\omega_{k}a_{4}(a_{3}p-q)+i\omega_{k}\sqrt{p^{2}+ 
r_{0}^{2}}\cos\theta}\no\\
&=&-\int_{-1}^{1}du\intsinf \frac{\omega_{k}d\omega_{k}}{8\pi^{2}} \intinf  
\frac{dp~e^{i\Delta{E}a_{1}p}}{2v_{A}v_{B}\gamma_{A}\gamma_{B}} \no\\
&&\times~ 
e^{-i\omega_{k}a_{3}a_{4}p+i\omega_{k}\sqrt{p^{2}+ r_{0}^{2}}u} 
\int_{-\infty}^{\infty}dq~e^{i(a_{2}\Delta{E}+a_{4}\omega_{k})q}\no\\&=&-
\int_{-1}^{1}du\intsinf \frac{\omega_{k}d\omega_{k}}{4\pi} \intinf  
\frac{dp~}{2v_{A}v_{B}\gamma_{A}\gamma_{B}}
e^{i\omega_{k}\sqrt{p^{2}+ r_{0}^{2}}u}\no\\&&\times~ e^{ip(a_{1}\Delta{E}-a_{3}a_{4}\omega_{k})} 
  ~\delta(a_{2}\Delta{E}+a_{4}\omega_{k})\no\\
 ~&=& 0~.
\end{eqnarray}

\section{Evaluations of the integrals in $(1+3)$ dimensions for 
detectors in perpendicular motion}\label{Appn:1p3-perpendicular-integrals}

\subsection{Explicit evaluation of $\mathcal{I}_{A}$}\label{Appn:1p3-perpendicular-integrals-IA}

We will evaluate $\mathcal{I}_{A}^{nv}$ for trajectory (\ref{eq:PerpTrajecA}). In particular, the 
$\mathcal{A}(\Delta E)$ part of this quantity becomes
\begin{eqnarray}
\mathcal{A}(\Delta E) &=& \intinf{d\tau_{A}}e^{i\Delta 
E\tau_{A}}\phi_{eff}(x_{A})\no\\
&=&\intinf{d\tau_{A}}e^{i\Delta 
E\tau_{A}}\intsinf\frac{\omega_{k} d\omega_{k}}{2(2\pi)^{3}}\int_{0}^{\pi}\sin\theta 
d\theta\int_{0}^{2\pi}d\phi\no\\
&&~~~~~~~~~~~~~~~\times e^{-i\omega_{k} t_{A}+i\omega_{k} v_{A} 
t_{A}\sin\theta\cos\phi}f(\omega_{k})\no\\
&=&
\intinf{d\tau_{A}}e^{i\Delta E\tau_{A}}\intsinf\frac{\omega_{k} 
d\omega_{k}}{2(2\pi)^{2}}\int_{-1}^{1}du~ 
e^{-i\omega_{k} t_{A}}f(\omega_{k})\no\\
&&~~~~~~~~~~~~~~~~~~~~~~~~~\times J_{0}(|\omega_{k} v_{A} t_{A}\sqrt{1-u^{2}}|)\no\\
&=& \intinf{\frac{dt_{A}}{\gamma}}e^{i\Delta Et_{A}/\gamma_{A}}\intsinf\frac{\omega_{k} 
d\omega_{k}}{2(2\pi)^{2}}e^{-i\omega_{k} 
t_{A}}f(\omega_{k}) \no\\~&&~~~~~~~~~~~~~~~~~~~~~~~~~~~~~\times~\frac{2\sin{\omega_{k} v_{A} 
t_{A}}}{\omega_{k} 
v_{A} t_{A}}\no\\
&=&
\intsinf\frac{ d\omega_{k}~f(\omega_{k})}{i2(2\pi)^{2}v_{A}\gamma_{A}}\intinf 
\frac{dt_{A}}{t_{A}}[e^{i(\Delta 
E-\frac{\omega_{k}}{D_{A}} )t_{A}/\gamma_{A}}\no\\
&&~~~~~~~~~~~~~~~~~~~~~~~~~ -e^{i(\Delta E-\omega_{k} 
D_{A})t_{A}/\gamma_{A}}]\no\\
&=&
\intsinf\frac{ d\omega_{k}}{i2(2\pi)^{2}v_{A}\gamma_{A}}f(\omega_{k})i\pi[\text{sgn}(D_{A}\Delta 
E-\omega_{k})\no\\
&&~~~~~~~~~~~~~~~~~~~~~~~~~-\text{sgn}(\Delta E/D_{A}-\omega_{k})]\no \\
&=&\frac{1}{4\pi v_{A}\gamma_{A}}\int_{\Delta E/D_{A}}^{\Delta E D_{A}}d\omega_{k}\, f(\omega_{k})~.
\end{eqnarray}
As all other quantities in $\mathcal{I}_{A}$ are zero, one can express $ \mathcal{I}_{A} = 
\left|\mathcal{A}(\Delta E)\right|^{2}$. This expression is same as the one from Eq. (\ref{IAgen}) 
for two detectors in parallel inertial motion in $(1+3)$ dimensions, as expected.

\subsection{Explicit evaluation of 
$\mathcal{I}_{\varepsilon}$}\label{Appn:1p3-perpendicular-integrals-Ie}

Now let us evaluate $\mathcal{I}_{\varepsilon_{W}}^{vac}$ for two detectors in perpendicular 
inertial motions. For $t_{A}>t_{B}$, we have positive frequency Wightman function
\begin{eqnarray}\label{1p3greenPerp2}
&G_{W}&(X_{A},X_{B})\no\\
&=& -\frac{1}{4\pi^{2}}\frac{1}{(t_{A}-t_{B}-i\epsilon)^{2}-|\boldsymbol{x}_{A}
-\boldsymbol{x}_{B}|^{2}}\no\\&=&-\frac{1}{4\pi^{2}}\frac{1}{g(t_{A},t_{B})-i\epsilon}~.
\end{eqnarray}
No contribution for $t_{B}>t_{A}$, will be taken. Now, solving for $g(t_{A},t_{B})-i\epsilon=0$ 
gives, $t_{B}=\gamma_{B}^{2}(t_{A}\pm u(t_{A})(1+i\epsilon))=t_{\pm}(1\pm i\epsilon)$. We already 
know that  $t_{B}=t_{+}>t_{A}$ and $t_{B}=t_{-}<t_{A}$. When $t_{A}>t_{B}$ satisfied, we have pole 
in the lower half of the complex $t_{B}$-plane. Thus 
\begin{eqnarray}
\mathcal{I}_{\varepsilon_{W}}^{vac} &=& - \intinf\intinf d\tau_{A} 
d\tau_{B}e^{i\Delta{E}(\tau_{A}+\tau_{B})}G_{W}(X_{A},X_{B})\no\\
~&=& 0~.
\end{eqnarray}

\end{appendix}

\bibliographystyle{apsrev}

\bibliography{bibtexfile}

\end{document}